\documentclass[iop,apj]{emulateapj}
\usepackage{natbib}
\usepackage{graphicx}
\usepackage{amssymb}
\usepackage{epsfig}
\usepackage{aasmacros}
\usepackage{apjfonts}
\def\kab{\hbox{$K_{\rm s}$}}

\def\ks{\hbox{$K_{\rm s}$}}
 \catcode`\@=11
\def\gsim{\ifmmode{\mathrel{\mathpalette\@versim>}}
    \else{$\mathrel{\mathpalette\@versim>}$}\fi}
\def\lsim{\ifmmode{\mathrel{\mathpalette\@versim<}}
    \else{$\mathrel{\mathpalette\@versim<}$}\fi}
\def\@versim#1#2{\lower 2.9truept \vbox{\baselineskip 0pt \lineskip 
    0.5truept \ialign{$\m@th#1\hfil##\hfil$\crcr#2\crcr\sim\crcr}}}
\catcode`\@=12
\shorttitle{COSMOS-WIRCam NEAR-INFRARED IMAGING SURVEY. I.}
\shortauthors{McCracken et al.}
\begin{document}
\submitted{Accepted for publication in The Astrophysical Journal}
\title {The COSMOS-WIRCam near-infrared imaging survey: I: $BzK$
  selected passive and star forming galaxy candidates at $z\gtrsim1.4$\altaffilmark{1}}
\altaffiltext{1}{Based on data collected at the Subaru Telescope,
which is operated by the National Astronomical Observatory of Japan;
the European Southern Observatory under Large Program 175.A-0839,
Chile and the Canada-France-Hawaii Telescope with WIRCam and
MegaPrime/MegaCam the latter operated as a joint project by the CFHT
Corporation, CEA/DAPNIA, the NRC and CADC of Canada, the CNRS of
France, TERAPIX and the University of Hawaii. This research has made
use of the NASA/IPAC Infrared Science Archive, which is operated by
the Jet Propulsion Laboratory (JPL), California Institute of Technology,
under contract with NASA. Support for this work was provided by the Spitzer Science Center which
is operated by JPL and Caltech under a contract with NASA.}
\author{
H.~J.~McCracken\altaffilmark{2},
P. Capak\altaffilmark{3,4}, 
M. Salvato\altaffilmark{3},
H. Aussel\altaffilmark{5},
D. Thompson\altaffilmark{6},
E. Daddi\altaffilmark{5},
D.B. Sanders\altaffilmark{7},
J.-P. Kneib\altaffilmark{8},
C.~J.  Willott\altaffilmark{9},
C. Mancini\altaffilmark{10}, 
A. Renzini\altaffilmark{10},
R. Cook\altaffilmark{11}, 
O. Le F\`evre\altaffilmark{8},
O. Ilbert \altaffilmark{7,8},
J. Kartaltepe \altaffilmark{8},
A.~M.~Koekemoer\altaffilmark{12},
Y. Mellier\altaffilmark{2},
T.~Murayama\altaffilmark{13}, 
N.Z. Scoville\altaffilmark{3},
Y.~Shioya\altaffilmark{13} and
Y.~Taniguchi\altaffilmark{13}
}
\altaffiltext{2}{Institut d'Astrophysique de Paris, UMR7095 CNRS, Universit\'{e} Pierre et Marie Curie, 98 bis Boulevard Arago, 75014 Paris, France}
\altaffiltext{3}{Spitzer Science Center, 314-6 Caltech, Pasadena, CA 9112, USA}
\altaffiltext{4}{California Institute of Technology, MS 104-25,
  Pasadena, CA, 91125, USA}
\altaffiltext{5}{Service d'Astrophysique, CEA/Saclay, 91191 Gif-sur-Yvette, France}
\altaffiltext{6}{Large Binocular Telescope Observatory, University of Arizona, 933 N. Cherry Ave.Tucson, AZ  85721-0065,  USA}
\altaffiltext{7}{Institute for Astronomy, University of Hawaii, 2680 Woodlawn Drive, Honolulu, HI, 96822, USA}
\altaffiltext{8}{Laboratoire d'Astrophysique de Marseille, CNRS- Universit\'e d'Aix-Marseille, 38 rue Fr\'ed\'eric Joliot-Curie, 13388 Marseille Cedex 13, France}
\altaffiltext{9}{Herzberg Institute of Astrophysics, National Research Council, 5071 West Saanich Road, Victoria, BC V9E 2E7, Canada}
\altaffiltext{10}{INAF, Osservatorio Astronomico Di Padova, Vicolo Osservatorio 2, I-35122, Padova, Italy}
\altaffiltext{11}{Department of Physics, Brown University, Box 1843,
  Providence, RI 02912, USA}
\altaffiltext{12}{Space Telescope Science Institute, 3700 San Martin
  Drive, Baltimore, MD 21218, USA}
\altaffiltext{13}{Research Center for Space and Cosmic Evolution, Ehime University, 2-5 Bunkyo-cho, Matsuyama 790-8577, Japan}

\begin{abstract}
  We present a new near-infrared survey covering the $2\deg^2$ COSMOS
  field conducted using WIRCam at the Canada--France--Hawaii
  Telescope. By combining our near-infrared data with Subaru $B$ and
  $z$ images, we construct a deep, wide-field optical-infrared
  catalogue. At $K_{\rm s}<23$ (AB magnitudes), our survey
  completeness is greater than $90\%$ and $70\%$ for stars and
  galaxies, respectively, and contains $143,466$ galaxies and $13,254$
  stars. Using the $BzK$ diagram, we divide our galaxy catalogue into
  quiescent and star-forming galaxy candidates. At $z\sim2$, our
  catalogues contain $3931$ quiescent and $25,757$ star-forming
  galaxies representing the largest and most secure sample at these
  depths and redshifts to date. Our counts of quiescent galaxies turns
  over at $K_{\rm s}\sim22$, an effect which we demonstrate cannot be
  due to sample incompleteness. Both the number of faint and bright
  quiescent objects in our catalogues exceeds the predictions of a
  recent semi-analytic model of galaxy formation, indicating
  potentially the need for further refinements in the amount of
  merging and active galactic nucleus feedback at $z\sim2$ in these
  models. We measure the angular correlation function for each sample
  and find that the slope of the field galaxy correlation function
  flattens to $1.5$ by $\ks\sim23$. At small angular scales, the
  angular correlation function for passive $BzK$ galaxies is
  considerably in excess of the clustering of dark matter.  We use
  precise 30-band photometric redshifts to derive the spatial
  correlation length and the redshift distributions for each object
  class. At $K_{\rm s}<22$ we find
  $r_0^{\gamma/1.8}=7.0\pm0.5h^{-1}$~Mpc for the passive $BzK$
  candidates and $4.7\pm0.8h^{-1}$~Mpc for the star-forming $BzK$
  galaxies. Our $pBzK$ galaxies have an average photometric redshift
  of $z_p\sim1.4$, in approximate agreement with the limited
  spectroscopic information currently available.The stacked $K_{\rm
    s}$ image will be made publicly available from IRSA.
\end{abstract}


\section{Introduction}
Understanding the formation and evolution of galaxies is one of the
central themes in observational cosmology. To address this issue
considerable work is underway to construct complete galaxy samples
over a broad redshift range using many diverse ground and space-based
facilities.  The ultimate aim of these studies is to construct
distribution functions of galaxy properties such as stellar mass, star
formation rate, morphology and nuclear activity as a function of
redshift and environment and to use these measurements in conjunction
with theoretical models to establish the evolutionary links from one
cosmic epoch to another. Essentially we would like to understand how
the galaxy population at distant times became the galaxies we see
around us today and to identify the physical mechanisms driving this
transformation. Improvements in both modelling and observations are the
ultimate drivers in achieving this understanding.

The cold dark matter (CDM) model of structure formation remains our
best description of how structures grow on large scales from minute
fluctuations in the cosmic microwave background
\citep{1996MNRAS.282..347M,Springel:2006p8432}. In this picture,
structures grow ``hierarchically'', with the smallest objects forming
first. Unfortunately, CDM only tells us about the underlying
collisionless dark component of the universe; what we observe are
luminous galaxies and stars. At large scales, the relationship between
dark matter and luminous objects (usually codified as the ``bias'') is
simple, but at small scales (less than a few megaparsecs) the effects
of baryonic physics intervene to make the relationship between
luminous and non-luminous components particularly
complex. Understanding how luminous objects ``light up'' in dense
haloes of dark matter one is essentially the problem of
understanding galaxy formation. ``Semi-analytic'' models avoid
computationally intensive hydrodynamic simulations by utilising a set
of scaling relations which connect dark matter to luminous objects (as
will be explained later in this paper), and these models are now
capable of predicting how host of galaxy properties, mass assembly and
star-formation rate evolve as a function of redshift.

Observationally, at redshifts less than one, various spectroscopic
surveys such as the VVDS
\citep{2005A&A...439..845L,2005A&A...439..863I,Pozzetti:2007p100}
DEEP2 \citep{Faber:2007p5801,Noeske:2007p5802} and zCSOSMOS-bright
\citep{Lilly:2007p1799,Silverman:2009p5804,Mignoli:2009p5805} have
mapped the evolution of galaxy and active galactic nuclei (AGNs)
populations over fairly wide areas on the sky. There is now general
agreement that star formation in the Universe peaks at $1<z<2$ and
that $\sim 50\%-70\%$ of mass assembly took place in the redshift
range $1<z<3$
\citep{Connolly:1997p8767,Dickinson:2003p8724,Arnouts:2007p3665,Pozzetti:2007p100,Noeske:2007p5802,PerezGonzalez:2008p8736}. Alternatively
stated, half of today's stellar mass appears to be in place by
$z\sim1$ \citep{Drory:2005p8438,Fontana:2004p8883}. This is largely at
odds with the predictions of hierarchical structure formation models
which have difficulty in accounting for the large number of evolved
systems at relatively early times in cosmic history
\citep{Fontana:2006p3228}. Furthermore, there is some evidence that
around half the stellar mass in evolved or ``passive'' galaxies
assembled relatively recently \citep{Bell:2004p8521}.  It is thus of
paramount importance to gather the largest sample of galaxies possible
at this redshift range.

In the redshift range $1.4 < z < 3.0 $ identifiable
spectral features move out of the optical wave bands and so
near-infrared imaging and spectroscopy become essential. The role of
environment and large-scale structure at these redshifts is largely
unexplored \citep{Renzini:2009p5799}. It is also worth mentioning that
in addition to making it possible to select galaxies in this important
range, near-infrared galaxy samples offer several advantages compared
to purely optical selections (see, for example
\cite{Cowie:1994p1284}). They allow us to select $z>1$ galaxies in
the rest-frame optical, correspond more closely to a
stellar-mass-selected sample and are less prone to dust extinction.
As $k-$ corrections in $K-$ band are insensitive to galaxy type over a
wide redshift range, near-infrared-selected samples provide a fairly
unbiased census of galaxy populations at high redshifts (providing
that the extinction is not too high, as in the case of some submillimeter
galaxies). Such samples represent the ideal input catalogues from
which to extract targets for spectroscopic surveys as well as for
determining accurate photometric redshifts.

\cite{Cowie:1996p8471} carried out one of the first extremely deep,
complete $K-$ selected surveys and made the important discovery that
star-forming galaxies at low redshifts have smaller masses than
actively star-forming galaxies at $z\sim1$, a phenomenon known as
``downsizing''. Stated another way, the sites of star-formation
``migrate'' from higher-mass systems at high redshift lower-mass
systems at lower redshifts.  More recent $K$-selected surveys include
the K20 survey \citep{Cimatti:2002p3013} reaching $\kab\simeq 21.8$
and the GDDS survey \citep{Abraham:2004p2980} which reached
$\kab\simeq 22.4$ provide further evidence for this
picture. The areas covered by these surveys was small,
comprising only $\sim 55$ arcmin$^2$ and $\sim 30$ arcmin$^2$ K20 and
GDDS respectively. While
\cite{Glazebrook:2004p3320} and \cite{Cimatti:2008p1800} provided spectroscopic
confirmation of evolved systems $z>1.4$ and provided further evidence
for the downsizing picture \citep{Juneau:2005p8688} their limited
coverage made them highly susceptible to the effects of cosmic
variance. It became increasingly clear that much larger samples of
passively evolving galaxies were necessary.

At $K<20$ the number of passive galaxies at $z\sim2$ redshifts is
small and spectroscopic followup of a complete magnitude-limited
sample can be time-consuming. For this reason a number of groups have
proposed and validated techniques based on applying cuts in
colour-colour space to isolate populations in certain redshift
ranges. Starting with the Lyman-break selection at $z\sim3$
\citep{Steidel:1996p9040}, similar techniques have been applied at
intermediate redshifts to select extremely red objects (EROs;
\cite{Hu:1994p9049}) or distant red galaxies (DRGs;
\cite{Franx:2003p310}) and the ``BzK'' technique used in this paper
\citep{Daddi:2004p76}. The advantage of these methods is that they are
easy to apply requiring at most only three or four photometric bands;
their disadvantage being that the relationships between each object
class is complicated and some selection classes contain galaxies with
a broad range of intrinsic properties
\citep{Daddi:2004p76,Lane:2007p295,Grazian:2007p398}. The relationship
to the underlying complete galaxy population can also be difficult to
interpret \citep{LeFevre:2005p8609}. Ideally, one like to make
complete mass-selected samples at a range of redshifts but such
calculations require coverage in many wave bands and can depend
sensitively on the template set
\citep{Pozzetti:2007p100,Longhetti:2009p9068}. Moreover, for redder
populations the mass uncertainties can be even larger;
\cite{Conroy:2009p9107} estimate errors as larger as 0.6 dex at $z\sim
2$.

At $z\sim1.4$ \cite{Daddi:2004p76} used spectroscopic data from the
K20 survey in combination with stellar evolutionary tracks to define
their ``BzK'' technique.  They demonstrated that in the $(B-z)$
$(z-K)$ colour-colour plane, star-forming galaxies and evolved systems
are well separated at $z>1.4$, making it possible accumulate larger
samples of passive galaxies at intermediate redshifts that was
possible previously with simple one-colour criterion.

Subsequently, several other surveys have applied these techniques to
larger samples of near-infrared selected galaxies. In one of the
widest surveys to date, \cite{Kong:2006p294} constructed $K$-band
selected samples over a $\sim 920$ arcmin$^2$ field reaching
$\kab\simeq 20.8$ reaching to $\kab\simeq 21.8$ over a 320 arcmin$^2$
sub-field. The exploration of a field of this size made possible to
measure the clustering properties of star-forming and passive galaxy
sample and to establish that passively evolving galaxies in this
redshift range are substantially more strongly clustered than
star-forming ones, indicating that a galaxy-type - density relation
reminiscent of the local morphology-density relation must be already
in place at $z\gsim 1.4$.

The UKIDSS survey reaches $\kab\sim22.5$ over a $\sim 0.62$-deg$^2$
area included in the Subaru-\textit{XMM Newton} Deep Survey and
\cite{Lane:2007p295} used this data set to investigate the different
commonly-used selected techniques at intermediate redshifts,
concluding most bright DRG galaxies have spectra energy distributions
consistent with dusty star-forming galaxies or AGNs at $\sim2$. They
observe a turn-over in the number counts of passive $BzK$ galaxies.

Other recent works include the MUSYC/ECDFS survey covering $\sim 900$
arcmin$^2$ to $\kab\sim 22.5$ over the CDF South field
\citep{Taylor:2008p3500}, not to be confused with the GOODS-MUSIC
catalog of \cite{Fontana:2006p3228}, which covers 160 arcmin$^2$ of
GOODS-South field to $\kab\sim 23.8$. This $K$-band selected
catalogue, as well as the FIREWORKS catalog by \cite{Wuyts:2008p5806},
are based on the ESO Imaging Survey coverage of the GOODS-South
field\footnote{http://www.eso.org/science/goods/releases/20050930./}
These studies have investigated, amongst other topics, the evolution
of the mass function at $z\sim2$ and what number of red sequence
galaxies which were already in place at $z\sim2$.

Finally, one should mention that measuring the distribution of a
``tracer'' population, either red passive galaxies or normal field
galaxies can provide useful additional information on the galaxy
formation process. In particular one can estimate the mass of the dark
matter haloes hosting the tracer population and, given a suitable
model for halo evolution, identify the present-day descendants of the
tracer population, as has been done for Lyman break galaxies at
$z\sim3$. A few studies have attempted this for passive galaxies at
$z\sim2$, but small fields of view have made these studies somewhat
sensitive to the effects of cosmic variance.  The ``COSMOS'' project
\citep{2007ApJS..172....1S} comprising a contiguous $2~\deg^2$
equatorial field with extensive multi-wavelength coverage, is well
suited to probing the universe at intermediate redshift.

In this paper we describe a $K_{\rm s}$-band survey covering the
entire $\sim 1.9\deg^2$ COSMOS field carried out with WIRCam at
the Canada-France-Hawaii Telescope (CFHT). The addition of deep, high
resolution $K-$ data to the COSMOS field enables us to address many of
scientific issues outlined in this introduction, in particular to
address the nature of the massive galaxy population in the redshift
range $1<z<2$.

Our principal aims in this paper are to (1) present a catalogue of
$BzK$-selected galaxies in the COSMOS field; (2) present the number
counts and clustering properties of this sample in order principally
to establish the catalogue reliability; (3) present the COSMOS $K-$
imaging data for the benefit of other papers which make indirect use
of this dataset (for example, in the computation of photometric
redshifts and stellar masses).

Several papers in preparation or in press make use of the data
presented here. Notably, \cite{Ilbert:2009p7351} combine this data
with IRAC and optical data to investigate the evolution of the galaxy
mass function. The deep part of the zCOSMOS survey
\citep{Lilly:2007p5770} is currently collecting large numbers galaxies
spectra at $z>1$ in the central part of the COSMOS field using a
colour-selection based on the $K$- band data set described here.

Throughout the paper we use a flat lambda cosmology ($\Omega_m~=~0.3$,
$\Omega_\Lambda~=~0.7$) with
$h~=~H_{\rm0}/100$~km~s$^{-1}$~Mpc$^{-1}$. All magnitudes are given in the
AB system, unless otherwise stated. The stacked $K_{\rm s}$ image
presented in this paper will made publicly available at
IRSA.\footnote{\url{http://irsa.ipac.caltech.edu/Missions/cosmos.html}}

\section{Observations and data reductions}
\label{sec:observ-data-reduct}

\subsection{Observations}
\label{sec:observations}

WIRCam \citep{Puget:2004p4595} is a wide-field near-infrared detector
at the 3.6m CFHT which consists of four $2048\times2048$ cryogenically
cooled HgCdTe arrays. The pixel scale is $0.3\arcsec$ giving a field
of view at prime focus of $21\arcmin\times21\arcmin$.  The data
described in this paper were taken in a series of observing runs
between 2005 and 2007. A list of these observations and the total
amount of on-sky integration time for each run can be found in
Table~\ref{tab:observations}.

Observing targets were arranged in a set of pointing centres arranged
in a grid across the COSMOS field. At each pointing centre
observations were shifted using the predefined ``DP10'' WIRCAM
dithering pattern, in which each observing cube of four micro-dithered
observations is offset by $1\arcmin 12\arcsec$ in RA and $18\arcsec$
in decl. Our overall observing grid was selected so as to fill all
gaps between detectors and provide a uniform exposure time per pixel
across the entire field and to ensure a the WIRCAM focal plane was
adequately sampled to aid in the computation of the astrometric
solution. All observations were carried out in queue-scheduled
observing mode. Our program constraints demanded a seeing better than
$0.8\arcsec$ and an air mass less than 1.2. Observations were only
validated by CFHT if these conditions were met. In practice, a few
validated images were outside these specifications (usually due to
short-term changes in observing conditions) , and were rejected at
later subsequent processing steps.

Observations were made in the $K_{\rm s}$ filter (``$K$-short'';
\citep{Skrutskie:2006p3517}), which has a bluer cutoff than the
standard $K$ filter, and unlike the $K'$ filter
\citep{1992AJ....103..332W} has a ``cut-on'' wavelength close to
standard $K$. This reduces the thermal background and for typical
galaxy spectra decreases the amount of observing time needed to reach
a given signal-to-noise ratio (S/N) compared to the standard $K-$
filter. A plot of the WIRCam $K_{\rm s}$ filter is available from
CFHT.\footnote{\url{http://www.cfht.hawaii.edu/Instruments/Filters/wircam.html}}

\subsection{Pre-reductions}
\label{sec:pre-reductions}

WIRCam images were pre-reduced using IRAF\footnote{\url
  {http://iraf.noao.edu/}} using a two-pass method. To reduce the data
volume the four micro-dithers were collapsed into a single frame at
the beginning of the reduction process.  This marginally reduced the
image quality (by less than $0.1\arcsec$ FWHM) but made the reduction
process manageable on a single computer. After stacking the sub-frames
the data were bias subtracted and flat fielded using the bias and flat
field frames provided by the CFHT WIRCam queue observing team.  A
global bad pixel mask was generated using the flat to identify the
dead pixels; the dark frames were used to identify hot pixels.  A
median sky was then created and subtracted from the images using all
images in a given dither pattern.  The images were then stacked using
integer pixel offsets and the WCS headers with IRAF's
\texttt{imcombine} task. These initial stacks of the science data are
used to generate relatively deep object masks through SExtractor's
'CHECKIMAGE\_TYPE = OBJECTS' output file.  These object masks are used
to explicitly mask objects when re-generating the sky in the second
pass reduction. Supplementary masks for individual images are made to
mask out satellites or other bad regions not included in the global
bad pixel mask on a frame-by-frame basis.

In the second pass reduction, we individually sky-subtract each
science frame using the object masked images and any residual
variations in the sky is removed by subtracting a constant to yield a
zero mean sky level.  These individual sky-subtracted images of a
single science frame are then averaged with both sigma-clipping to
remove cosmic rays and masking using a combination of the object mask
and any supplementary mask to remove the real sources and bad regions
in the sky frames.  The region around the on-chip guide star is also
masked. These images are further cleaned of any non-constant residual
gradients as needed by fitting to the fully masked (object $+$
supplementary $+$ global bad pixel masks) background on a line-by-line
basis. Any frames with poor sky subtraction or other artifacts after
this step were rejected.
	
Next amplifier cross talk is removed for bright Two Micron All Sky
Survey (2MASS) stars which creates ``donut'' shaped ghosts at
regularly spaced intervals. These are unfortunately variable, with
their shape and level depending on the brightness of the star and the
amplifier the star falls on. We proceed in three steps: we first build
for each star a median image of the potential cross talk pattern it
could generate. This is done by taking the median of $13\times13$
pixels sub-frames at 64 pixel intervals above and below the star
position, taking into account the full bad pixel mask. Next, we check
whether the cross-talk is significant by comparing the level of the
median image to the expected noise. If cross-talk is detected, we
subtract it by fitting at every position it could occur the median
shape determined in the preceding steps.

After pre-reduction, the TERAPIX tool \texttt{QualityFITS} was used to
create weight-maps, catalogues, and quality assessment Web pages for
each image. These quality assessment pages provide information on the
instrument point-spread function (PSF), galaxy counts, star counts,
background maps and a host of other information. Using this
information, images with focus or electronic problems were
rejected. \texttt{QualityFITS} also produces a weight-map for each
WIRCam image; this weight map is computed from the bad pixel mask and
the image flat-field. Observations with seeing FWHM greater than
$1.1\arcsec$ were rejected.

\begin{deluxetable}{ccc}
\tablewidth{0pt}
\tabletypesize{\scriptsize}
\tablecolumns{3}
\small
\tabletypesize{\scriptsize}
\tablecaption{COSMOS-WIRCam observations\label{tab:observations}}
\tablewidth{0pt}
\tablehead{
\colhead{Year} & \colhead{RunID} &\colhead{total integration time (hrs)}
}
\startdata
2005 &BH36 & 8.1\\
2005 &BH89 & 3.5\\	
2006 &AH97 & 12.8\\	
2006 &AC99 & 2.1\\
2006 &BF97 & 12.7\\	
2006 &BH22 & 13.3\\	
2007 &AF34 & 6\\	
2007 &AC20 & 6.1\\	
2007 &AH34 & 17\\	
\enddata
\end{deluxetable}

\subsection{Astrometric and photometric solutions}
\label{sec:astr-solut}

In the next processing step the TERAPIX software tool \texttt{Scamp}
\citep{2006ASPC..351..112B} was used to compute a global astrometric
solution using the COSMOS i* catalogue \cite{Capak:2007p267} as
an astrometric reference. 

{\texttt {Scamp} calculates a global astrometric solution. For each
astrometric "context" (defined by the QRUNID image keyword supplied
by CFHT) we derive a two-dimensional third-order polynomial which
minimises the weighted quadratic sum of differences in positions
between the overlapping sources and the COSMOS reference astrometric
catalogue derived from interferometric radio observations. Each
observation with the same context has the same astrometric
solution. Note that we use the {\texttt STABILITY\_MODE}
``instrument'' parameter setting which assumes that the derived
polynomial terms are identical exposure to exposure within a given
context but allows for anamorphosis induced by atmospheric
refraction. This provides reliable, robust astrometric solutions for
large numbers of images. The relative positions of each wircam
detector are supplied by precomputed header file which allows us to
use the initial, approximate astrometric solution supplied by CFHT
as a first guess.

Our internal and external astrometric accuracies are $\sim0.2$ and
$\sim0.1$ arcseconds respectively. Scamp produces an XML summary file
containing all details of the astrometric solution, and any image
showing a large reduced chisquared is flagged and rejected.

\begin{figure}
\resizebox{\hsize}{!}{\includegraphics{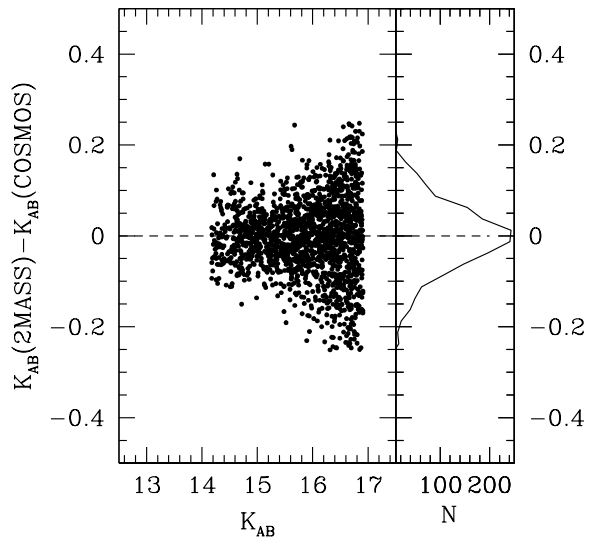}}
\caption{The difference between COSMOS-WIRCam total magnitudes and
  2MASS magnitudes as a function of COSMOS-WIRCam total magnitude.}
\label{fig:2mass_comparison}
\end{figure}

We do not use \texttt{Scamp} to compute our photometric solutions (the
version we used (1.2.11MP) assumes that the relative gains between
each WIRCam detector is fixed). Instead, we first use the astrometric
solution computed by \texttt{scamp} to match 2MASS stars with objects
in each WIRCam image. We then compute the zero-point of each WIRCam
detector by comparing the fluxes of 2MASS sources with the ones
measured by \texttt{SExtractor} in a two-pass process. First, saturated objects
brighter than $K_{\rm AB}=13.84$ magnitudes or objects where the
combined photometric error between SExtractor and 2MASS is greater
than 0.2 magnitudes are rejected. An initial estimate of the zero-point is 
produced by computing the median of the difference between the two
cleaned catalogues. Any object where this initial estimation differ by
more than $3\sigma$ from the median is rejected, and the final
difference is computed using an error-weighted mean. The difference
between 2MASS and COSMOS catalogues in the stacked image is shown in
Figure~\ref{fig:2mass_comparison}. Note that the main source of
scatter at these magnitudes ($13.84 < K_{\rm AB} < 17$) comes from
uncertainties in 2MASS photometry. The magnitude range with good
sources between objects in 2MASS and WIRCam catalogues is quite narrow
(around two magnitudes) so setting these parameters is quite
important.

Finally, all images and weight-maps were combined using \texttt{Swarp}
\citep{Bertin:2002p5282}.  The tangent point used in this paper was
$10^{\mathrm h} 00^{\mathrm m} 15^{\mathrm s}$,$+02\degr 17\arcmin
34.6\arcsec$ (J2000) with a pixel scale of 0.15''/pixel to match
COSMOS observations in other filters. The final image has an effective
exposure time of one second and a zero point of $31.40$ AB
magnitudes. This data will be publically available from the IRSA web
site. The seeing on the final stack is excellent, around $0.7\arcsec$
FWHM. Thanks to rigorous seeing constraints imposed during
queue-scheduled observations, seeing variation over the final stack is
small, less than $\sim5\%$.

\subsection{Complementary datasets}
\label{sec:compl-datas}

We also add Subaru-suprime $B_J$, $i^+$ and $z^+$ imaging data
(following the notation in \cite{Capak:2007p267}). We downloaded the
image tiles from IRSA\footnote{\url{http://irsa.ipac.caltech.edu/}}
and recombined them with \texttt{Swarp} to produce a
single large image astrometrically matched to the $K_{\rm s}-$band WIRCam
image (the astrometric solutions for the images at IRSA were
calculated using the same astrometric reference catalogue as our
current $K_{\rm s}-$ image, and they share the same tangent point). Catalogues
were extracted using \texttt{SExtractor} \citep{1996A&AS..117..393B}
in dual-image mode, using the $K_{\rm s}-$band image as a detection image. An
additional complication arises from the fact that the $B-$ Subaru
images saturate at $B\sim19$. To account for this, we use the TERAPIX
tool \texttt{Weightwatcher} to create ``flag-map'' images in which all
saturated pixels are indicated (the saturation limit was determined
interactively by examining bright stars in the images).  During the
subsequent scientific analysis, all objects which have flagged pixels
are discarded. We also manually masked all bright objects by defining
polygon region files. In addition, we also automatically masked
regions at fixed intervals from each bright star to remove positive
crosstalk in the $K_{\rm s}-$ image. The final catalogue covers a total area
of $1.9\deg^2$ after masking.

\section{Catalogue preparation and photometric calibration}
\label{sec:catal-prep}

\subsection{Computing  colours}
\label{sec:comp-total-colo}

We used \texttt{SExtractor} in dual-image mode with the $K_{\rm
  s}-$band as a reference image to extract our catalogues. For this
$K_{\rm s}-$selected catalogue, we measured $K_{\rm s}-$ band total
magnitudes using \texttt{SExtractor}'s \texttt{MAG\_AUTO}
  measurement and aperture colours.  Our aperture magnitudes are
measured in a diameter of $2\arcsec$ and we compute a correction to
``total'' magnitudes by comparing the flux of point-like sources in
this small aperture with measurements in a larger $6\arcsec$ diameter
aperture. We verified that for the $B_J$ and $z^+$ Subaru-suprime
images the difference between these apertures varies less than $0.05$
magnitudes, indicating that seeing variations are small across the
images which was confirmed by an analysis of the variation of the
best-fit Gaussian FWHM for $B,z$ and $K$ images over the full
$2\deg^2$ field.

  Obviously for extended bright objects this colour measurement will
be dominated the object nucleus as the majority of the $z\sim2$
objects studied in this paper are unresolved, distant galaxies we will
neglect this effect. We verified that for these objects,
variable-aperture colours computed using \texttt{MAG\_AUTO} gave
results very similar (within 0.1~mag) to these corrected aperture
colours.

Based on these considerations, we apply the following corrections to
our aperture measurements to compute colours:

\begin{equation}
i^+_{\rm tot} = i^+-0.1375
\end{equation}

\begin{equation}
K_{\rm tot} = K'-0.1568
\end{equation}

\begin{equation}
B_{\rm tot} = B_J - 0.1093
\end{equation}

For the $z^+$ our corrections are more involved. 

As noted in \cite{Capak:2007p267} the Subaru $z$-band images were
taken over several nights with variable seeing. To mitigate the
effects of seeing variation on the stacked image PSF individual
exposures were smoothed to the same (worst) FWHM with a Gaussian
before image combination.  This works well at faint magnitudes where
many exposures were taken so the non-Gaussian wings of the PSF average
out.  However, at bright magnitudes ($z^+\sim20$) the majority of
longer exposures are saturated, so the non-Gaussian wings of the PSF
in the few remaining exposures can bias aperture photometry. To
correct for this non-linear effect we apply a magnitude dependent
aperture correction in the transition magnitudes between
$19<z<20$. After first applying the correction to total magnitude,

\begin{equation}
z^+{_{\rm tot}} = z^+ - 0.1093
\end{equation}

We apply a further correction, for $z^+<19.0$, $z^+_{\rm
  tot}=z^+_{\rm tot}-0.023$ and for $z^+>20.0$, $z^+_{\rm tot}=z^+_{\rm tot}+0.1$. For
$19<z^+<20$,
\begin{equation}
z^+_{\rm tot} = z^+_{\rm tot}+(z^+_{\rm tot}-19.0) \times 0.077+0.023
\end{equation}

Flux errors in \texttt{SExtractor} are underestimated (in part due to
correlated noise in the stacked images) and must be corrected. For a
given $4000\times4000$ image section, we compute a correction factor
from the ratio of the one-sigma error between a series of $2\arcsec$
apertures on empty regions of the sky and the median
\texttt{SExtractor} errors for all objects. The mean correction factor
is derived from several such regions. For $B$,$i$ and $z$ images we
multiply our flux errors by 1.5; for the $K_{\rm s}$ image we apply a
correction factor of $2.0$. 

\subsection{Catalogue completeness and limiting magnitude}
\label{sec:catal-compl-limit}

\begin{figure}
\resizebox{\hsize}{!}{\includegraphics{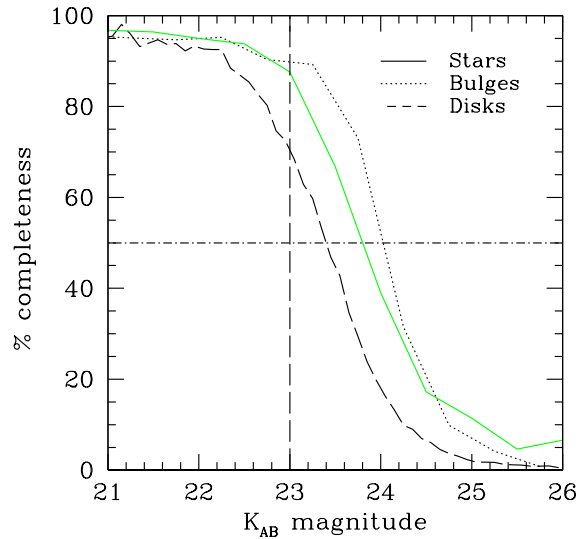}}
\caption{Recovery fraction for point-like sources, bulges and disks
for the central region of the COSMOS-WIRCam stack. Note that the
slight decline in completeness at relatively bright magnitudes ($K_{
\rm {AB}} \sim 22$) is due to confusion.}
\label{fig:compsim}
\end{figure}

We conducted an extensive set of realistic simulations to determine
limiting magnitude as a function of object magnitude and profile. In
our simulations, we first created a noiseless image containing a
realistic mix of stars, disk-dominated and bulge-dominated galaxies
using the TERAPIX software \texttt{stuff} and
\texttt{skymaker}. Type-dependent luminosity functions are
  their evolution with redshift are taken from the VVDS survey
  \citep{Zucca:2006p7671,Ilbert:2006p3432}. The spectral energy
  distribution (SED) of each galaxy type was modelled using empirical
  templates of \cite{1980ApJS...43..393C}. The disk size distribution
  was modelled using the fitting formula and parameters presented in
  \cite{deJong:2000p7685}.

  Next, we used \texttt{SExtractor} to detect all objects on the
  stacked image (using the same configuration used to detect objects
  for the real catalogues) and to produce a \texttt{CHECK\-IMAGE} in
  which all these objects were removed (keyword
  \texttt{CHECKIMAGE\_TYPE -OBJECTS}). In the next step this empty
  background was added to the simulated image, and \texttt{SExtractor}
  run again in ``assoc-mode'' in which a match is attempted between
  each detected galaxy and the output simulated galaxy catalogue
  produced by \texttt{stuff}. In the last step, the magnitude
  histograms of the number output galaxies is compared to the
  magnitude histograms in the input catalogue; this ratio gives the
  completeness function of each type of object. In total
    around 30,000 objects (galaxy and stars) in one image were used in
    these simulations. The results from these simulations are shown
  in Figure~\ref{fig:compsim}. The solid line shows the completeness
  curve for stars and the dotted line for disks. The completeness
  fraction is $70\%$ for disks and $90\%$ for stars and bulges at
  $K_{\rm s}\sim23$.

In addition to these simulations, we compute upper limits for each
filter based on simple noise statistics in apertures of $2\arcsec$
(after applying the noise correction factors listed above). The
$1\sigma$, $2\arcsec$ limiting magnitude for our data at the centre of
the field are $29.1$,$27.0$ and $25.4$~AB magnitudes for $B$, $z$ and
$K_{\rm s}$ magnitudes respectively. 

A related issue is the uniformity of the limiting magnitude over the
full image. Figure~\ref{fig:depthmap} shows the limiting magnitude as
a function of position for the $K_{\rm s}$ stack. This was created by
converting the weight map to an rms error map, scaling this error map
from $1\sigma$ per pixel to units of $5\sigma$ in the effective area
of an optimally weighted $0.7\arcsec$ aperture, and then converting
this flux to units of AB magnitude. This depth map agrees well with
the completeness limit for point sources shown
Figure~\ref{fig:compsim}. Note that future COSMOS-WIRCam observations
(which will be available in around one years' time)  are expected to
further reduce the depth variations across the survey area. 

\begin{figure}
\plotone{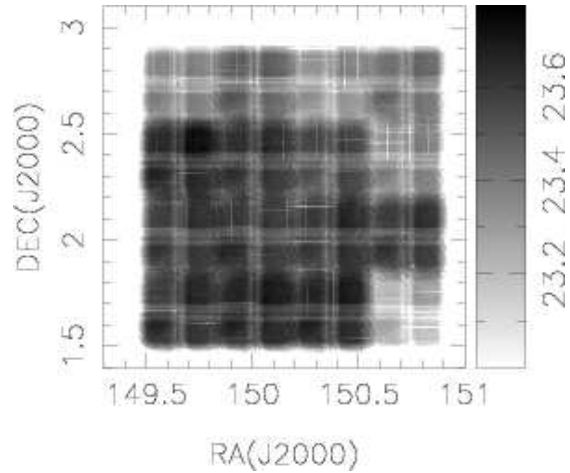}
\caption{Depth map for the COSMOS-WIRCam survey. This map was
  constructed from the weight-map from the final stacked image. The
  grey-scale corresponds to the magnitude at which a point source is
  detected at $5\sigma$ in a $2\arcsec$ aperture.} 
\label{fig:depthmap}
\end{figure}

Based on the considerations outlined in this section, we adopt $K_{\rm
  s}=23$ as the limiting magnitude for our catalogues. At this limit
our catalogue is greater than $90\%$ and $70\%$ complete for
point-sources and galaxies respectively. At this magnitude the number
of spurious sources (based on carrying out detections on an image of
the $K_s$ stack multiplied by -1) is less than $1\%$ of the total.
\subsection{The $BzK$ selection}
\label{sec:bzk-selection}

\begin{figure}
\resizebox{\hsize}{!}{\includegraphics{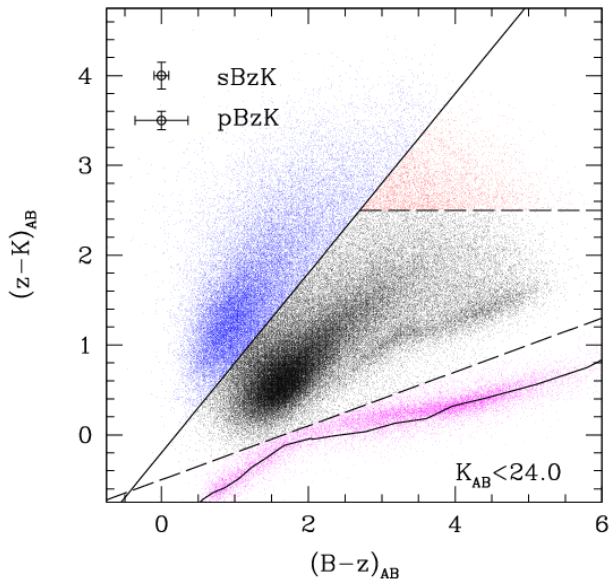}}
\caption{The $(B-Z)_{\rm AB}$ vs $(z-K)_{\rm AB}$ diagram for all
  galaxies in the COSMOS field. Four distinct regions are shown: stars
  (lower part of the diagram), galaxies (middle), star-forming
  galaxies (left) and passively-evolving galaxies (top right). The solid
  line shows the colours of stars in the $BzK$ filter set of
  \citeauthor{Daddi:2004p76} computed using the models of
  \cite{Lejeune:1997p4534}} 
\label{fig:bzk_diagram}
\end{figure}

One of the principal objectives of this paper is to produce a
reliable catalogue of objects at $z\sim2$ using a colour-colour
selection technique. A number of different methods now exist to select
galaxies in colour-colour space. For instance, the ``dropout''
technique \citep{Steidel:1996p9040} makes use of Lyman-break
spectral feature and the opacity of the high-redshift universe to
ultraviolet photons to select star-forming galaxies at $z\sim3$,
provided they are not too heavily reddened. Similar techniques can be
used at $1<z<3$ \citep{Adelberger:2004p9303,Erb:2003p9269} and large
samples of UV-selected star-forming galaxies now exist at these
redshifts. At intermediate redshifts, spectroscopy has shown that
``ERO'' galaxies which are galaxies selected according to red
optical-infrared colours contains a mix of old passive galaxies and
dusty star-forming systems in the redshift range $0.8<z<2$
\citep{Cimatti:2002p3013,Yan:2004p9408}. The ``DRG'' criteria, which
selects galaxies with $(J-K)_{\rm Vega}>2.3$ is affected by similar
difficulties \citep{Papovich:2006p9480,Kriek:2006p7151}.  On the other
hand, the ``$BzK$'' criterion introduced by \cite{Daddi:2004p76} can
reliably select galaxies in the redshift range $1.4\lsim z\lsim 2.5$
with relatively high completeness and low contamination. Based on the
location of star-forming and reddened systems in a spectroscopic
control sample lie in the $(B-z)$ and $(z-K)$ plane and considerations
of galaxy evolutionary tracks, it has been adopted and tested in
several subsequent studies (e.g.,
\cite{Kong:2006p294,Lane:2007p295,Hayashi:2007p3726,Blanc:2008p3635,Dunne:2009p6812,Popesso:2009p6632}). $BzK-$selected
galaxies are estimated to have masses of $\sim 10^{11}M_\odot$ at
$z\sim2$ \citep{Daddi:2004p76,Kong:2006p294}.

Compared to other colour criteria, it offers the advantage of
distinguishing between actively star-forming and passively-evolving
galaxies at intermediate redshifts. It also sharply separates stars
from galaxies, and is especially efficient for $z>1.4$ galaxies. The
criterion was originally designed using the redshift evolution in the
$BzK$ diagram of various star-forming and passively-evolving template
galaxies (i.e., synthetic stellar populations) located over a wide
redshift interval. \citeauthor{Daddi:2004p76} carried out extensive
verifications of their selection criteria using spectroscopic
redshifts.  

To make the comparison possible with previous studies we wanted our
photometric selection criterion to match as closely as possible as the
original ``BzK'' selection proposed in \citeauthor{Daddi:2004p76} and
adopted by the authors cited above. As our filter set is not the same
as this work we applied small offsets (based on the tracks of
synthetic stars), following a similar procedure outlined in
\cite{Kong:2006p294}.

To account for the differences between our Subaru $B-$ filter and the
$B-$ VLT filter used by \cite{Daddi:2004p76} we use this
empirically-derived transformation, defining $bz=B_{J_{\rm
    total}}-z^+_{\rm tot}$, then for blue objects with $bz<2.5$,

\begin{equation}
bz_{\rm cosmos}=bz+0.0833\times bz+0.053 
\end{equation}

otherwise, for objects with $bz>2.5$, 

\begin{equation}
bz_{\rm cosmos}= bz + 0.27
\end{equation}

This ``$bz_{\rm cosmos}$'' quantity is the actual corrected
$(B-z)_{\rm AB}$
colour which we use in this paper. 

Finally we divide our catalogue into galaxies at $z<1.4$, stars,
star-forming galaxies and passively evolving galaxies at $1.4<z<2.5$,
by first defining the $BzK$ quantity introduced in
\cite{Daddi:2004p76}:

\begin{equation}
 BzK\equiv(z-K)-(B-z)
\end{equation}

For galaxies expected at $z>1.4$ star-forming galaxies (hereafter
$sBzK$) are selected as those objects with $BzK>-0.2$. One
should also note that the reddening vector in the $BzK$ plane is
approximately parallel to the $sBzK$ selection criteria, which ensure
that the selection is not biased against heavily reddened dusty galaxies.

Old, passively
evolving galaxies (hereafter $pBzK$) can be selected as those objects
which have

\begin{equation}
BzK<-0.2, (z-K)>2.5. 
\end{equation}  

Stars are selected using this criteria:

\begin{equation}
(z-K) < -0.5+(B-z) \times 0.3
\end{equation}  

Finally, the full galaxy sample consists simply of objects which do
\textit{not} fulfill this stellarity criterion. The result of this
division is illustrated in Figure~\ref{fig:bzk_diagram}. The solid
line represents the colours of stars in the $BzK$ filter set of
\citeauthor{Daddi:2004p76} using the empirically corrected spectra
presented in \cite{Lejeune:1997p4534}, and it agrees very with our
corrected stellar locus. 

\section{Source counts}
\label{sec:source-number-counts}

We now present number counts of the three populations selected in the
previous Section.  

\subsection{Star and galaxy counts}
\label{sec:galaxy-counts}

\begin{figure}
\resizebox{\hsize}{!}{\includegraphics{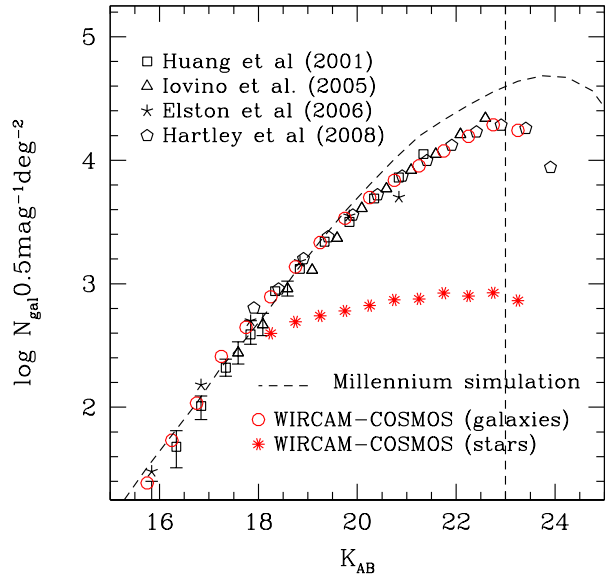}}
\caption{$K_{\rm s}-$ selected galaxy and star counts from the COSMOS survey
  (open circles and stars respectively) compared to measurements from
  recent wide-field near-infrared surveys.}
\label{fig:counts_all}
\end{figure}

\begin{deluxetable*}{ccccccc}
\tabletypesize{\scriptsize}
  \tablecolumns{7}
\tablewidth{0pt}
\small
    \tablecaption{Differential Number Counts\\ of Stars, Galaxies and
      Star-Forming galaxies per Half-magnitude Bin.\label{tab:countsgals}}
\tablehead{
&\multicolumn{2}{c}{Galaxies} & 
\multicolumn{2}{c}{Stars} &\multicolumn{2}{c}{$sBzK's$} \\
\colhead{$K_{\rm AB}$}&\colhead{$N_{\rm gal}$}&\colhead{$\log(N_{\rm gal})$~$\deg^{-2}$}&\colhead{$N_{\rm stars}$}&\colhead{$\log(N_{\rm stars})$~$\deg^{-2}$}&\colhead{$N_{\mathit sBzK}$}&\colhead{$\log(N_{\mathit sBzK})$~$\deg^{-2}$}}
\startdata
16.25&       102& 1.73&\nodata &\nodata &\nodata&\nodata\\
16.75&       204& 2.03&\nodata&\nodata  &\nodata&\nodata\\
17.25&       487& 2.41&\nodata & \nodata&\nodata&\nodata\\
17.75&       838& 2.65&\nodata & \nodata&\nodata&\nodata\\
18.25&      1479& 2.89&        750& 2.60   &\nodata&\nodata\\
18.75&      2588& 3.14&        928& 2.69   &\nodata&\nodata\\
19.25&      4073& 3.33&       1038& 2.74&        24& 1.10\\
19.75&      6410& 3.53&       1138& 2.78 &      61& 1.51\\
20.25&      9433& 3.70&       1257& 2.82&       195& 2.01\\
20.75&     12987& 3.84&       1397& 2.87&       710& 2.57\\
21.25&     17027& 3.95&       1425& 2.88&      1982& 3.02\\
21.75&     22453& 4.07&       1586& 2.92&      4191& 3.35\\
22.25&     29502& 4.19&       1504& 2.90&      7684& 3.61\\
22.75&     36623& 4.29&       1596& 2.93&     11109& 3.77\\
\enddata
\tablecomments{{Note. Logarithmic counts are normalised to the effective area
of our survey, $1.89\deg^2$.}} 
\end{deluxetable*}
Figure~\ref{fig:counts_all} shows our differential galaxy number
counts compared to a selection of measurements from the literature. We
note that at intermediate magnitudes ($20<K_s<22$) counts from the
four surveys presented here are remarkably consistent
\citep{Elston:2006p1644,1997ApJ...476...12H,Hartley:2008p5290}. At
$16<K_s<20$ discrepancies between different groups concerning
measurement of total magnitudes and star-galaxy separation leads to an
increased scatter. At these magnitudes, shot noise and
large-scale-structure begin to dominate the number count errors.

The COSMOS-WIRCam survey is currently the only work to provide
unbroken coverage over the range $16 < K_s<23$. In addition, our
colour-selected star-galaxy separation provides a very robust way to
reject stars from our faint galaxy sample. These stellar counts are
shown by the asterisks in Figure~\ref{fig:counts_all}. We note that at
magnitudes brighter than $K_{\rm s}\sim18.0$ our stellar number counts
become incomplete because of saturation in the Subaru $B$ image (our
catalogues exclude any objects with saturated pixels which
preferentially affect point-like sources). Our galaxy and star number
counts are reported in Table~\ref{tab:countsgals}.

\subsection{$sBzK$ and $pBzK$ counts}
\label{sec:sbzk-pbzk-counts}

\begin{figure}[htb!]
\resizebox{\hsize}{!}{\includegraphics{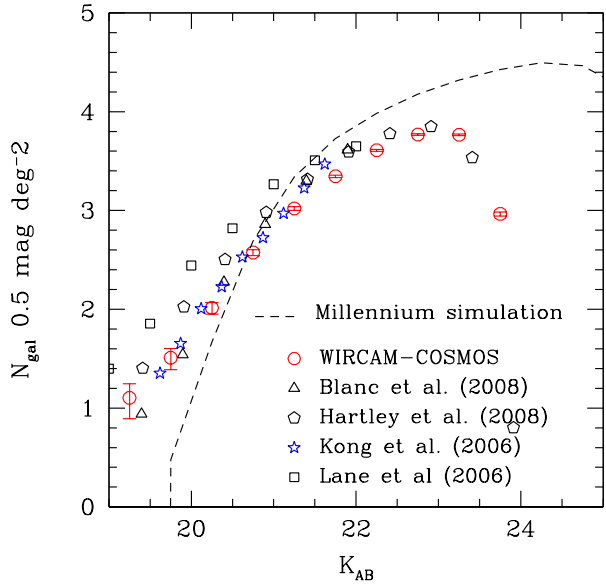}}
\caption{Number counts for star-forming $BzK$ galaxies in the
  COSMOS-WIRCam survey (open circles) compared to measurements from
  the literature and the predictions of the model of
  \citeauthor{Kitzbichler:2007p3449} (dashed line).}
\label{fig:counts_sbzk}
\end{figure}

Figure~\ref{fig:counts_sbzk} shows the counts of star-forming $BzK$
galaxies compared to measurements from the literature. These counts
are summarised in Table~\ref{tab:countsgals}. We note an excellent
agreement with the counts in \cite{Kong:2006p294} and the counts
presented by the MUYSC collaboration \citep{Blanc:2008p3635}. However,
the counts presented by the UKIDSS-UDS group
\citep{Lane:2007p295,Hartley:2008p5290} are significantly offset
compared to our counts at bright magnitudes, and become consistent
with it by $K_{\rm s}\sim 22$. These authors attribute the
discrepancy to cosmic variance but we find photometric offsets a more
likely explanation (see below).

Figure~\ref{fig:pbzk_newsel} shows in more detail the zone occupied by
passive galaxies in Figure~\ref{fig:bzk_diagram}. Left of the diagonal
line are objects classified as star-forming $BzK$ galaxies. Objects
not detected in $B$ are plotted as right-pointing arrows with colours
computed from the upper limit of their $B-$ magnitudes. An
  object is considered undetected if the flux in a 2\arcsec aperture
  is less than the corrected $1\sigma$ noise limit. For the $B-$ band
  this corresponds to approximately 29.1 mag. This criterion
means that in addition to the galaxies already in the $pBzK$ selection
box, fainter $sBzK$ with $B$-band non-detections (shown with the green
arrows) may be scattered rightward into the $pBzK$ region.

\begin{figure}
\resizebox{\hsize}{!}{\includegraphics{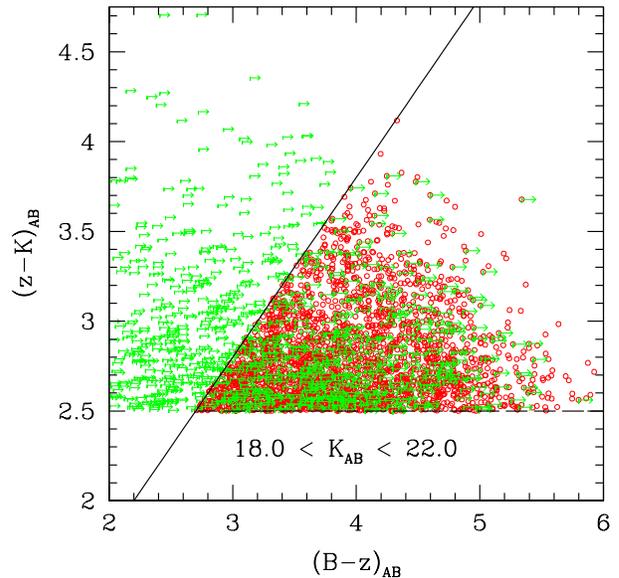}}
\caption{Selection diagram for the passive $BzK$ population. Objects
  with rightward-pointing arrows are galaxies plotted at the lower
  limit of their $(B-z)_{\rm AB}$ colours. Circles are objects
  selected as $pBzK$ galaxies; normal $sBzK$ galaxies are not shown.}
\label{fig:pbzk_newsel}
\end{figure}

Counts for our passive galaxy population including these
``additional'' objects are represented by the hatched region in
Figure~\ref{fig:counts_pbzk}. The upper limit for the source counts in
this figure represents the case in which \textit{all} the $(z-K)_{s} >
2.5$ sources undetected in $B$ are scattered into the $pBzK$
region. Even accounting for these additional objects we unambiguously
observe a flattening and subsequent turnover in the passive galaxy
counts at around $K_{\rm s}\sim22$, well above the completeness limit of
either our $K_{\rm s}-$ or $B-$ data in agreement with
\citep{Hartley:2008p5290}. 

This upper limit, however, is a conservative estimate. We have made a
better estimate of this upper limit by carrying out a stacking
analysis of the objects not detected in $B-$ in both the passive and
star-forming regions of the $BzK$ diagram. For each apparent $K_{\rm s}$
magnitude bin in Table we median-combine Subaru $B-$band postage
stamps for objects with no $B$-band detection, producing separate
stacks for the star-forming and passive regions of the $BzK$
diagram. In both cases, objects below our detection limit are clearly
visible (better than a three-sigma detection) in our stacked images at
each magnitude bin to $K_{\rm s}~23$. By assuming that the mean $B$
magnitude of the stacked source to be the average magnitude of our
undetected sources, we can compute the average $(B-z)$ colour of our
undetected sources, and reassign their location in the $BzK$ diagram
if necessary. This experiment shows that at most only $15\%$ of the
star-forming $BzK$ galaxies undetected in $B-$ move to the passive
$BzK$ region.

Our number counts are summarised in Table~\ref{tab:countspbzk}, which
also indicates the upper count limits based on $B-$ band
observations. As before, our counts are in good agreement with those
presented in \cite{Kong:2006p294} and \cite{Blanc:2008p3635} but
are above the counts in \cite{Hartley:2008p5290}.

To investigate the origin of this discrepancy, we compared our $BzK$
diagram with \citeauthor{Hartley:2008p5290}'s, which should also be in
the \citeauthor{Daddi:2004p76} filter set. We superposed our $BzK$
diagram on that of \citeauthor{Hartley:2008p5290} and found that the
\citeauthor{Hartley:2008p5290} stellar locus is bluer by $\sim0.1$ in
both $(B-z)$ and $(z-K)$ compared to our measurements\footnote{Since
  the first draft of this manuscript was prepared, a communication
  with W. Hartley has confirmed that the transformations to the Daddi
  et al. system were incorrectly computed in their work.}. We have
already seen that our stellar locus agrees well with the theoretical
stellar sequence computed using the \cite{Lejeune:1997p4534} synthetic
spectra in the \citeauthor{Daddi:2004p76} filter set and also with the
stellar locus presented in \citeauthor{Daddi:2004p76} and
\citeauthor{Kong:2006p294}. We conclude therefore that the number
counts discrepancies arise from an incorrect transformation to the
\citeauthor{Daddi:2004p76} filter set.

\begin{figure}
\resizebox{\hsize}{!}{\includegraphics{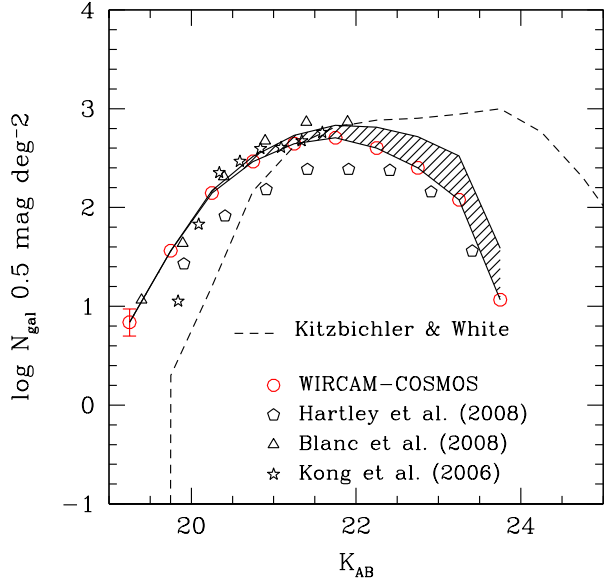}}
\caption {Differential number counts for the passive $BzK$ population
  in the COSMOS-WIRCam survey (open circles) compared to measurements
  from the literature and the predictions of the model of
  \citeauthor{Kitzbichler:2007p3449} (dashed line). The shaded
  region represents an upper limit on the number counts of passive
  $BzKs$ if all star-forming BzKs in Figure~\ref{fig:pbzk_newsel} were
  moved into the region of the figure occupied by the passively
  evolving population. }
\label{fig:counts_pbzk}
\end{figure}

\begin{deluxetable}{ccccc}
\tabletypesize{\scriptsize}
\tablewidth{0pt}
  \tablecaption{Differential Number Counts for the passive $BzK$ population.\label{tab:countspbzk}}
\tablehead{
    &\multicolumn{2}{c}{{Passive $BzK$}}
    &\multicolumn{2}{c}{{{Passive $BzK$ (upper limits)}}\tablenotemark{a}} \\
\colhead{$K_{\rm AB}$} & \colhead{$N_{\mathit pBzK}$}& \colhead{$\log(N_{\mathit pBzK})$~$\deg^{-2}$} &
\colhead{$N_{\mathit pBzK}$} & \colhead{$\log(N_{\mathit pBzK})$~$\deg^{-2}$}}
\startdata
19.25&13&0.84&13&0.84\\
19.75&69&1.56&69&1.56\\
20.25&265&2.15&280&2.17\\
20.75&553&2.47&621&2.52\\
21.25&837&2.65&1015&2.73\\
21.75&963&2.71&1285&2.83\\
22.25&757&2.60&1229&2.81\\
22.75&475&2.40&984&2.72\\
\enddata
\tablecomments{{Logarithmic counts are normalised to the effective area
of our survey, $1.89\deg^2$.}} 
\tablenotetext{a}{The upper limit to the $pBzK$ was computed by including
  all the $sBzK$ galaxies undetected in $B$ with $(z-K)>2.5$}
\end{deluxetable}

\subsection{Comparison with the Semi-analytic Model of~\citeauthor{Kitzbichler:2007p3449}}
\label{sec:comparison-with-semi}

In Figures \ref{fig:counts_all},~\ref{fig:counts_sbzk}
and~\ref{fig:counts_pbzk} we show counts of galaxies extracted from
the semi-analytical model presented in
\cite{Kitzbichler:2007p3449}. 

Semi-analytic models start from either an analytic ``merger tree'' of
dark matter haloes or, in the case of the model used here, merger
trees derived from a numerical simulation, the Millennium simulation
\citep{2005Natur.435..629S}). Galaxies are ``painted'' onto dark
matter haloes using a variety of analytical recipes which include
treatments of gas cooling, star-formation, supernovae feedback, and
black hole growth by accretion and merging. An important recent
advance has been the addition of ``radio mode'' AGN feedback
\cite{Bower:2006p7511,2007MNRAS.374.1303C} which helps provide a
better fit to observed galaxy luminosity functions. The \citeauthor
{Kitzbichler:2007p3449} model is derived from the work presented by
\cite{Croton:2006p7487} and further refined by
\cite{DeLucia:2007p7510}. It differs only from these papers in the
inclusion of a refined dust model. We refer the readers to these works
for further details. An extensive review of semi-analytic modelling
techniques can be found in \cite{Baugh:2006p782}.

To derive counts of quiescent galaxies, we follow the approach of
\cite{Daddi:2007p2924} and select all galaxies at $z>1.4$ in the
star-formation rate - mass plane (Figure 18 from
\citeauthor{Daddi:2007p2924}) which have star-formation rates less
than three times the median value for a given mass. Star-forming
objects were defined as those galaxies which do \texttt{not} obey
this criterion, in this redshift range. (Unfortunately the
  publicly available data do not contain all the COSMOS bands so we
  cannot directly apply the $BzK$ selection criterion to them.).

  In all three plots, the models over-predict the number of faint
  galaxies, an effect already observed for the $K-$ selected samples
  investigated in \citeauthor{Kitzbichler:2007p3449}. We also
    note that adding an upper redshift cut to the model catalogues to
    match our photometric redshift distributions (see later) does not
    change appreciably the number of predicted galaxies.

Considering in more detail the counts of quiescent galaxies we find
that at $20<\kab<20.5$ models are below observations by a factor of
two, whereas at $22.5<\kab<23.0$ model counts are in excess of
observations by around a factor of 1.5. Given the narrow redshift
range of our passive galaxy population, apparent \ks magnitude is a
good proxy for absolute \ks~magnitude which can itself be directly
related to underlying stellar mass \cite{Daddi:2004p76}. This implies
that these models predict too many small, low-mass passively-evolving
galaxies and too few large high mass passively evolving galaxies at
$z\sim1.4$. 

It is instructive to compare our results with Figure 7 from
\citeauthor{Kitzbichler:2007p3449}, which shows the stellar mass
function for their models. At $z\sim2$, the models both under-predict
the number of massive objects and over-predict the number of less
massive objects, an effect mirroring the overabundance of luminous
$pBzK$ objects with respect to the \citeauthor{Kitzbichler:2007p3449}
model seen in our data. 

A similar conclusion was drawn by \cite{Fontana:2006p3228} who
recently compared predictions for the galaxy stellar mass function for
massive galaxies for a variety of models with observations of massive
galaxies up to $z\sim4$ in the GOODS field. They also concluded that
models incorporating AGN feedback similar to
\citeauthor{Kitzbichler:2007p3449} under-predicted the number of high
mass galaxies.

Thanks to the wide-area, deep $B-$ band data available in the COSMOS
field, we are able to make reliable measurements of the number of
faint passive $BzK$ galaxies. Reassuringly, the turnover in counts of
passive galaxies observed in our data is qualitatively in agreement
with the measurements of the faint end of the mass function of
quiescent galaxies at $1.5<z<2$ made in \cite{Ilbert:2009p7351}.

\section{Photometric redshifts for the $pBzK$ and $sBzK$ population}
\label{sec:phot-redsh-pbzk}

For many years studies of galaxy clustering at $z\sim2$ have been
hindered by our imperfect knowledge of the source redshift
distribution and small survey fields. Coverage of the $2\deg^2$ COSMOS
field in thirty broad, intermediate and narrow photometric bands has
enabled the computation of very precise photometric redshifts
\citep{Ilbert:2009p4457}. 

These photometric redshifts were computed using deep Subaru data
described in \cite{Capak:2007p267} combined with intermediate band
data, the $K_{\rm s}$ data presented in this paper, $J$ data from
near-infrared camera WFCAM at the United Kingdom Infrared Telescope
and IRAC data from the Spitzer-COSMOS survey (sCOSMOS,
\cite{Sanders:2007p3108}). These near- and mid-infrared bandpasses
are an essential ingredient to compute accurate photometric redshifts
in the redshift range $1.4<z<2.5$, in particular because they permit
the location of the $4000$~\AA~break to be determined
accurately. Moreover, spectroscopic redshifts of 148 $sBzK$ galaxies
with a $\bar z\sim 2.2$ from the early zCOSMOS survey
\citep{Lilly:2007p1799} have been used to check and train these
photometric redshifts in this important redshift range.

  A set of templates generated by \cite{Polletta:2007p6857} using the
GRASIL code \cite{Silva:1998p6890} are used . The nine galaxy
templates of \citeauthor{Polletta:2007p6857} include three SEDs of
elliptical galaxies and six spiral galaxies templates (S0, Sa, Sb, Sc,
Sd, Sdm). This library is complemented with 12 additional blue
templates generated using the models of\cite{Bruzual:2003p963}.

  The photometric redshifts are computed using a standard $\chi^2$
template fitting procedure (using the ``Le Phare'' code). Biases in
the photo-z are removed by iterative calibration of the photometric
band zero-points. This calibration is based on 4148 spectroscopic
redshifts at $i^+_{AB}<22.5$ from the zCOSMOS survey
\cite{Lilly:2007p1799}. As suggested by the data, two different dust
extinction laws \cite{Prevot:1984p6814,Calzetti:2000p6839} were
applied specific to the different SED templates. A new method to
account for emission lines was implemented using relations between the
UV continuum and the emission line fluxes associated with star
formation activity.

  Based on a comparison between photometric redshifts and 4148
spectroscopic redshifts from zCOSMOS, we estimate an accuracy of
$\sigma_{\Delta z/(1+z)}=0.007$ for the galaxies brighter than
$i^+_{\rm AB}=22.5$. We extrapolate this result at fainter magnitude based
on the analysis of the 1$\sigma$ errors on the photo-z. At $z<1.25$,
we estimate an accuracy of $\sigma_z=0.02$, $\sigma_z=0.07$ at
$i^+_{\rm AB}\sim 24$, $i^+_{\rm AB}<25.5$, respectively.

\subsection{Photometric redshift distributions}

\label{sec:phot-redsh-distr}
\begin{figure}\resizebox{\hsize}{!}{\includegraphics{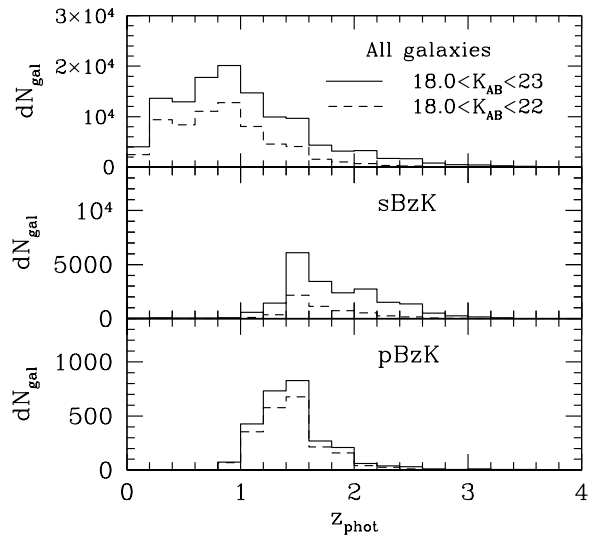}}
\caption{The Redshift distribution for field galaxies (top
panel) $sBzK$ (middle panel) and $pBzK$ galaxies (bottom panel),
computed using the 30-band photometric redshifts presented in
\cite{Ilbert:2009p4457}.}
\label{fig:zeddist}
\end{figure}

 We have cross-correlated our catalogue with photometric redshifts to
derive redshift selection functions for each photometrically-defined
galaxy population. Note that although photometric redshifts are based
on an optically-selected catalogue, this catalogue is very deep
($i'<26.5$) and contains almost all the objects present in the $K_{\rm
s}$-band selected catalogue. At $K_{\rm s}<23.0$, 138376 were
successfully assigned photometric redshifts, representing $96\%$ of
the total galaxy population.

Figure~\ref{fig:zeddist} shows the redshift distribution for all
$K_{\rm s}$-selected galaxies, as well as for $BzK$-selected
passively-evolving and star-forming galaxies in the magnitude range
$18.0<\kab<23.0$. We have computed the redshift selection function in
several magnitude bins and found that the effective redshift $z_{\rm
  eff}$ does not depend significantly on apparent magnitude for the
$sBzK$ and $pBzK$ populations.

By using only the blue grism of the VIMOS spectrograph at the VLT, the
zCOSMOS-Deep survey is not designed to target $pBzK$ galaxies, and so
no spectroscopic redshifts were available to train the photometric
redshifts of objects over the COSMOS field. At these redshifts the
main spectral features of $pBzK$ galaxies, namely Ca II H\& K and the
4000 \AA\ break, have moved to the near infrared. Hence, optical
spectroscopy can only deliver redshifts based on identifying the so
called Mg-UV feature at around 2800\AA~ in the rest frame. All in all,
spectroscopic redshifts of passive galaxies at $z>1.4$ are now
available for only a few dozen objects
\citep{Glazebrook:2004p3320,Daddi:2005p68,Kriek:2006p7151,McGrath:2007p7049,Cimatti:2008p1800}. We note
that the average spectroscopic redshift of these objects ($\bar
z\sim1.7$) indicates that the average photometric redshift of $\bar z
\sim 1.4$ of our $pBzK$ galaxies to the same $K_{\rm s}$-band limit may be
systematically underestimated. For the medium to short term one has to
unavoidably photometric redshifts when redshifts are needed for large
numbers of passive galaxies at $z>1.4$. 

\begin{figure}
\resizebox{\hsize}{!}{\includegraphics{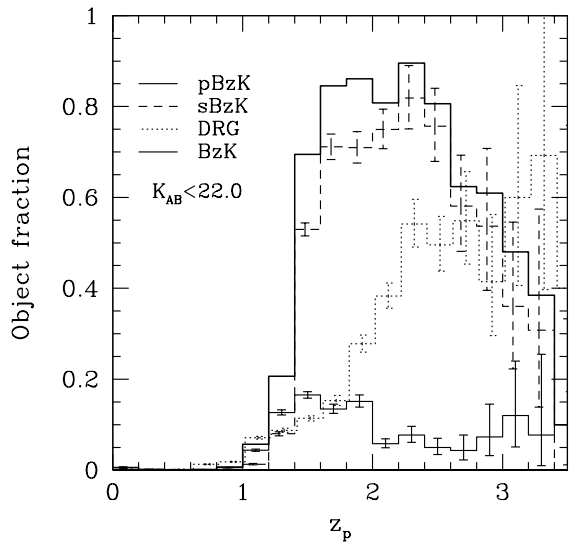}}
\caption{Fraction of the total galaxy population of DRG, $pBzK$ and
  $sBzK$ galaxy classifications. Poisson error bars in each bin have
  been offset slightly for clarity. The thick solid line represents the sum
  of $pBzK$ and $sBzK$ components.}
\label{fig:rel_dndz}
\end{figure}

From Figure~\ref{fig:zeddist} we can estimate the relative
  contribution of each classification type to the total number of
  galaxies to at least $z\sim2$. This is shown in
  Figure~\ref{fig:rel_dndz}, where we show for each bin in photometric
  redshift the number of each selection class as a fraction of the
  total number of galaxies. Upper and lower confidence limits are
  computed using Poisson statistics and the small-number approximation
  of \cite{Gehrels:1986p1797}; in general we can make reliable
  measurements to $z\sim2$. In the redshift range $1<z<3$ at
  $\ks\sim22$ the $pBzK$ population represents around $\sim20\%$ of
  the total number of galaxies, in contrast to $\sim70\%$ for
  $sBzK$-selected galaxies. The sum of both components represents at
  most $\sim 80\%$ of the total population at $z\sim2$.

We estimate the fraction of DRG galaxies using $J-$band data described
  in \cite{Capak:2007p267}. DRG-selected galaxies remain an
  important fraction of the total galaxy population, reaching around
  $\sim 50\% $ of the total at $z\sim2$. This in contrast with
  \cite{Reddy:2005p2062} who found no significant overlap in the
  spectroscopic redshift distributions of $pBzK$ and DRG galaxies.

Our work confirms that \textit{most} bright passive-$BzK$
  galaxies lie in a narrower redshift range than either the $sBzK$ or
  the DRG selection. We also see that the distribution in photometric
  redshifts for the DRG galaxies is quite broad. Similar conclusions
  were reached by \cite{Grazian:2007p398} and \cite{Daddi:2004p76}
  using a much smaller, fainter sample of galaxies.

\section{Clustering properties}
\label{sec:clust-prop-full}

\subsection{Methods}
\label{sec:meth}

For each object class we measure $w$, the angular correlation
function, using the standard \citet{1993ApJ...412...64L} estimator:

\begin{equation}
w ( \theta) ={\mbox{DD} - 2\mbox{DR} + \mbox{RR}\over \mbox{RR}}
\label{eq:1.ls}
\end{equation}

where $DD$, $DR$ and $RR$ are the number of data--data, data--random
and random--random pairs with separations between $\theta$ and
$\theta+\delta\theta$.  These pair counts are appropriately
normalised; we typically generate random catalogues with ten times
higher numbers of random points than input galaxies. We compute $w$ at
a range of angular separations in logarithmically spaced bins from
$\log(\theta)=-3.2$ to $\log(\theta)=-0.2$ with
$\delta\log(\theta)=0.2$, where $\theta$ is in degrees. At each
angular bin we use bootstrap errors to estimate the errors in
$w$. Although these are not in general a perfect substitute for a full
estimate of cosmic variance (e.g. using an ensemble of numerical
simulations), they should give the correct magnitude of the
uncertainty \citep{1992ApJ...392..452M}. 

We use a sorted linked list estimator to minimise the computation time
required. The fitted amplitudes quoted in this paper assume a
power-law slope for the galaxy correlation function,
$w(\theta)=A_w\theta^{1-\gamma}$; however this amplitude
must be adjusted for the `integral constraint' correction, arising
from the need to estimate the mean galaxy density from the sample
itself. This can be estimated as \citep[e.g.][]{2005ApJ...619..697A},
\begin{equation}
C = {1 \over {\Omega^2}} \int\!\!\! \int w(\theta)\, d\Omega_1\, d\Omega_2,
\label{eq:5}
\end{equation}

Our quoted fitted amplitudes are are corrected for this integral
constraint, i.e., we fit 

\begin{equation}
w(\theta)=A_w(\theta^{1-\gamma}-C) 
\end {equation}

For the COSMOS field, $C=1.42$ for $\gamma=1.8$. An added complication
is that the integral constraint correction depends weakly on the
slope,$\gamma$; in fitting simultaneously $\gamma$ and $A_w$ we use an
interpolated look-up table of values for $C$ in our minimisation
procedure.

Finally, it should be mentioned that in recent years it has become
increasingly clear that the power-law approximation for $w(\theta)$ is
no longer appropriate (see for example, \cite{Zehavi:2004p856}). In
reality, the observed $w(\theta)$ is the sum of the contributions of
galaxy pairs in separate dark matter haloes and within the same halo
of dark matter; it is only in a few fortuitous circumstances that this
observed $w(\theta)$ is well approximated by a power law of slope
$\gamma=1.8$. We defer a detailed investigation of the shape of
$w(\theta)$ in terms of these ``halo occupation models'' to a second
paper, but these points should be borne in mind in the forthcoming
analysis.

\subsection{Clustering of galaxies and stars}
\label{sec:galaxies-stars}
To verify the stability and homogeneity of our photometric calibration
over the full $2~\deg^2$ of the COSMOS field we first compute
the correlation function for stellar sources. These stars, primarily
residing in the galactic halo, should be
unclustered. They are identified as objects below the diagonal line in
Figure~\ref{fig:bzk_diagram}. This classification technique is more
robust than the usual size or compactness criterion, which can include
unresolved galaxies.

\begin{figure}[htb!]
\resizebox{\hsize}{!}{\includegraphics{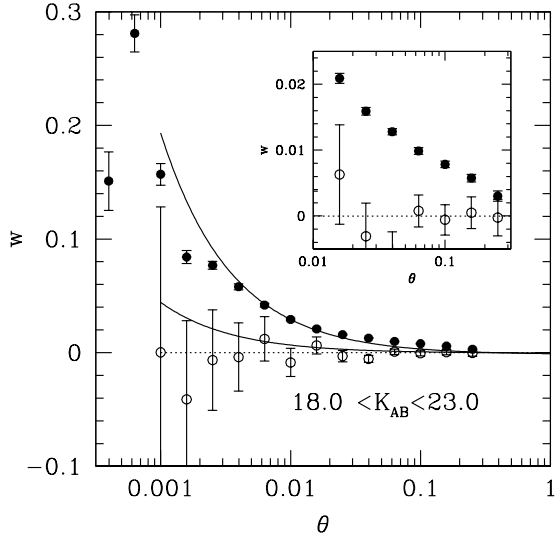}}
\caption{Clustering amplitude for stars and galaxies (open and filled
  circles respectively) in our catalogue selected with
  $18.0<K_{\rm AB}<23.0$ as a function of angular separation $\theta$. The
  inset shows measurements at large scales. The clustering amplitude
  of stars is consistent with zero at all angular scales.}
\label{fig:w_stars}
\end{figure}

The amplitude of $w(\theta)$ as a function of angular scale for stars
and faint galaxies is shown in Figure~\ref{fig:w_stars}. For
comparison we have also plotted the clustering amplitude for our
faintest $K_{\rm s}-$selected galaxy sample. The inset plot shows a
zoom on measurements at large scale scales where the amplitude of $w$
is very low. At each angular bin our stellar correlation function is
consistent with zero out to degree scales down to a limiting magnitude
of $K_{\rm s}=23$. If we fit a power-law correlation function of slope
0.8 to our stellar clustering measurements we find $A_w=(1.7\pm
1.7)\times 10^{-4}$ (at $1^\circ$; in comparison, the faintest
galaxy correlation function signal we measure is
$A_w=(9.9\pm1.5)\times10^{-4}$, around $\sim6$ times larger.

Figure~\ref{fig:w_slices} shows $w(\theta)$ for galaxies
in three magnitude slices. It is clear that the slope
of $w$ becomes shallower at fainter magnitudes. At small
separations (less than $1\arcsec$) $w$ decreases due to object
blending. Our fitted correlation amplitudes and slopes for field
galaxies are reported in Table~\ref{tab:fitw}.  

\begin{deluxetable}{ccc}
\tablecaption{Angular Correlation Amplitudes\label{tab:fitw}}
\tablehead{
\colhead{$K_{\rm AB}$} &\colhead{$A_w(1\arcmin)\times10^{-2}$} & \colhead{$\gamma$}
}

\tablewidth{0pt}
\startdata
 18.5 & $16.70\pm 4.03$ &$ 1.75\pm 0.06$\\
 19.0 & $12.10\pm 2.39$ &$ 1.74\pm 0.05$\\
 19.5 & $ 9.86\pm 1.43$ &$ 1.76\pm 0.03$\\
 20.0 & $ 7.83\pm 0.98$ &$ 1.72\pm 0.03$\\
 20.5 & $ 6.77\pm 0.70$ &$ 1.67\pm 0.02$\\
 21.0 & $ 5.69\pm 0.54$ &$ 1.61\pm 0.02$\\
 21.5 & $ 4.71\pm 0.42$ &$ 1.59\pm 0.02$\\
 22.0 & $ 3.81\pm 0.32$ &$ 1.59\pm 0.02$\\
 22.5 & $ 3.10\pm 0.26$ &$ 1.59\pm 0.02$\\
 23.0 & $ 2.57\pm 0.20$ &$ 1.55\pm 0.02$\\
\enddata
\end{deluxetable}

\begin{figure}
\resizebox{\hsize}{!}{\includegraphics{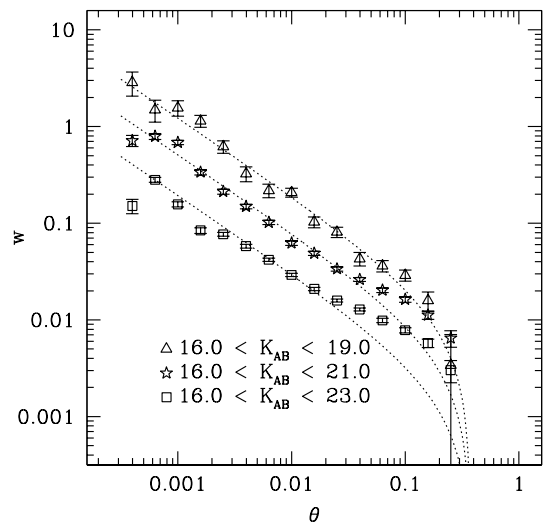}}
\caption{Clustering amplitude $w$ for galaxies in three slices of
  apparent magnitude. The dotted line shows a fit to a slope
  $\gamma=1.8$ with an integral constraint appropriate to the size of
  our field applied.}
\label{fig:w_slices}
\end{figure}

\begin{figure}
\resizebox{\hsize}{!}{\includegraphics{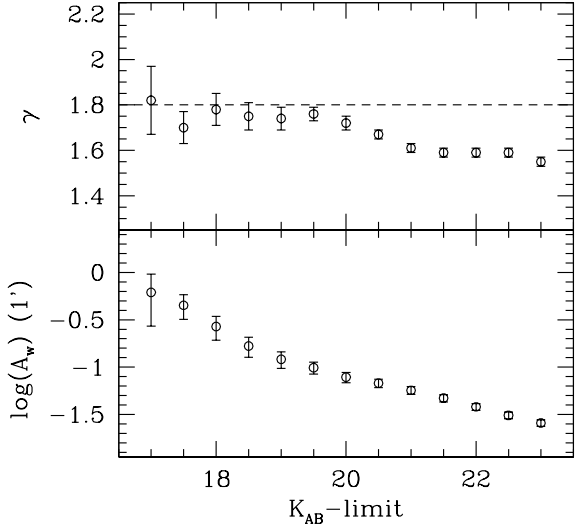}}
\caption{Lower panel: clustering amplitude at 1\arcmin~as a function
  of $K_{\rm s}$ limiting magnitude for the full galaxy sample. Upper
  panel: Best-fitting slope over entire angular range of our survey
  ($-3.2<\log(\theta)<0.2)$. }
\label{fig:scaling_w}
\end{figure}

In Figure~\ref{fig:scaling_w} we investigate further the dependence of
slope $\gamma$ on $K_{\rm s}$ limiting magnitude. Here we fit for the slope
and amplitude simultaneously for all slices. At bright magnitudes the
slope corresponds to the canonical value of $\sim 1.8$; towards
intermediate magnitudes it becomes steeper and fainter magnitudes
progressively flatter. It is interesting to compare this Figure with
the COSMOS optical correlation function presented in Figure 3 of
\cite{2007ApJS..172..314M} which also showed that the slope of the
angular correlation function becomes progressively shallower at
fainter magnitudes. One possible interpretation of this behaviour is
that at bright magnitudes our $K_{\rm s}$-selected samples are dominated by
bright, red galaxies which have an intrinsically steeper correlation
function slope; our fainter samples are predominantly bluer,
intrinsically fainter objects with shallower intrinsic correlation
function slope.

Finally, it is instructive to compare our field galaxy clustering
amplitudes with literature measurements as our survey is by far the
largest at these magnitude limits. Figure~\ref{fig:scaling_w_comp}
shows the scaling of the correlation amplitude at one degree as a
function to limiting $K_{\rm s}$ magnitude, compared a compilation of
measurements from the literature. To make this comparison, we have
assumed a fixed slope of $\gamma=1.8$ and converted the limiting
magnitude of each of our catalogues to Vega magnitudes. 

In general our results are within the $1\sigma$
error bars of most measurements, although it does appear that the
COSMOS field is slightly more clustered than other fields in the
literature, as we have discussed previously \citep{2007ApJS..172..314M}.

\begin{figure}[htb!]
\resizebox{\hsize}{!}{\includegraphics{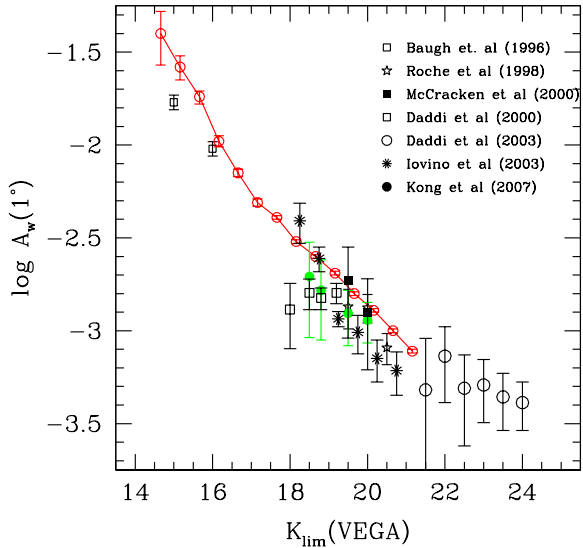}}
\caption{Fitted clustering amplitude at 1 degree as a function of
  $K_{\rm VEGA}$
  limiting magnitude (connected open circles), compared to values from
  the literature.}
\label{fig:scaling_w_comp}
\end{figure}

\nocite{1996MNRAS.283L..15B}
\nocite{Iovino:2005p17}
\nocite{1998MNRAS.295..946R}
\nocite{Daddi:2000p677}
\subsection{Galaxy clustering at $z\gtrsim 1.4$}
\label{sec:galaxy-clustering-at}

In the previous Sections we have demonstrated the reliability of our
estimates of $w$ and our general agreement with preceding
literature measurements for magnitude-limited samples. We now
investigate the clustering properties of passive and star-forming
galaxy candidates at $z\sim2$ selected using our $BzK$
diagram. Figure~\ref{fig:xyplot} shows the spatial distribution of the
$pBzK$ galaxies in our sample; a large amount of small-scale
clustering is evident.

\begin{figure}[htb!]
\resizebox{\hsize}{!}{\includegraphics{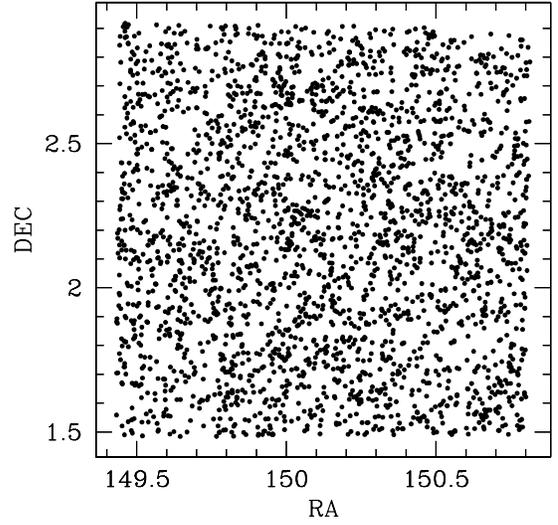}}
\caption{Angular distribution of~$18<K_{\rm s}<23$~$pBzK$ sources in the
  COSMOS-WIRCam survey. A large amount of small-scale clustering
  is clearly visible.}
\label{fig:xyplot}
\end{figure}

\begin{deluxetable}{ccccc}
\tablecaption{Angular correlation  amplitudes.\label{tab:fitw_pbzksbzk}}
\tablewidth{0pt}
\tablehead{
   &\multicolumn{2}{c}{Passive $BzK$}
    &\multicolumn{2}{c}{Star-forming $BzK$} \\
\colhead{$K_{\rm s}$} &\colhead{$A_w(1\arcmin)\times10^{-2}$} & \colhead{$\gamma$}
&\colhead{$A_w(1\arcmin)\times10^{-2}$} & \colhead{$\gamma$}}
\tablewidth{0pt}
\startdata
 22.0 & $ 8.41\pm 4.15$ &$ 2.32\pm 0.10$& $ 5.62\pm 1.72$ &$ 1.80\pm 0.07$\\
 23.0 & $ 6.23\pm 3.06$ &$ 2.50\pm 0.09$& $ 3.37\pm 0.62$ &$ 1.80\pm 0.04$\\
\enddata
\end{deluxetable}

The upper panel of Figure~\ref{fig:pbzk_w} shows the angular correlation functions
for our $pBzK$, $sBzK$ and for all galaxies. In each case we apply a
$18.0<K_s<23.0$ magnitude cut. For comparison we show the
clustering amplitude of dark matter computed using the redshift
selection functions presented in Section~\ref{sec:phot-redsh-pbzk} and
the non-linear power spectrum  approximation given in
\cite{2003MNRAS.341.1311S}.  At
intermediate to large scales, the clustering amplitude of field
galaxies and the $sBzK$ population follows very well the underlying
dark matter.  

The lower panel of Figure~\ref{fig:pbzk_w} shows the bias $b$, as a
function of scale, computed simply as $b(\theta)=\sqrt{(w_{\rm
    gal}(\theta)/w_{\rm dm}(\theta)}$. Dashed, dotted and solid lines show
$b$ values for $pBzK$, $sBzK$ and field galaxies retrospectively (in
this case our $w$ measurements have been corrected for the integral
constraint). The bias for the faint field galaxy population is
1.2 at $1\arcmin$ indicating that the faint
$K_{\rm s}-$ selected galaxy population traces well the underlying
dark matter. In comparison, at the same scales, the bias values for the
passive $BzK$ and star-forming $BzK$ galaxies are 2.5 and 2.1 respectively.

Our best-fitting $\gamma$ and amplitudes (quoted at $1\arcmin$) for
$pBzK$ and $sBzK$ galaxies are reported in
Table~\ref{tab:fitw_pbzksbzk}. Given that for the $sBzK$ galaxies
$\gamma=1.8$ we may compare with previous authors who generally assume
a fixed slope $\gamma=1.8$ for all measurements. At $K_{\rm VEGA}<20$,
corresponding to $K_{\rm AB}\sim22$, \cite{Kong:2006p294} find
$(4.95\pm0.52)\times10^{-3}$ whereas (at $1\deg$) we measure
$(2.1\pm0.6)\times10^{-3}$, closer to the value of
$(3.14\pm1.12)\times 10^{-3}$ found by \cite{Blanc:2008p3635}. We note
that both \cite{Hayashi:2007p3726} and \cite{2005ApJ...619..697A} also
investigated the luminosity dependence of galaxy clustering at
$z\sim$ although with samples considerable smaller than those
presented here. It is plausible that field-to-field variation and
large scale structure are the cause of the discrepancy between these surveys. 

The best fitting slopes for our $pBzK$ populations is $\gamma
\sim2.3$, considerably steeper than the field galaxy population (no
previous works have attempted to fit both slope and amplitude
simultaneously for the $pBzK$ populations due to small sample
sizes). In the next section we will derive the spatial clustering
properties of both populations.

\begin{figure}[htb!]
\resizebox{\hsize}{!}{\includegraphics{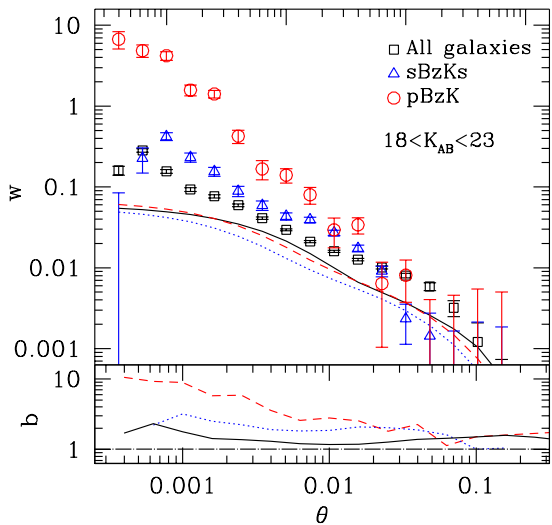}}
\caption{Top panel: amplitude of the galaxy correlation function $w$ for
  field galaxies, star-forming $BzK$ galaxies and passive $BzK$
  galaxies with $18<K_{\rm sAB}<23$ (squares, triangles and circles). The
  lines show the predictions for the non-linear clustering amplitudes
  of dark matter computed using the non-linear power spectrum. Bottom
  panel: bias, $b$ for $pBzK$, $sBzK$ and field galaxies (dashed,
  dotted and solid lines respectively).}
\label{fig:pbzk_w}
\end{figure}

\subsection{Spatial clustering}
\label{sec:spatial-clustering}

To de-project our measured clustering amplitudes and calculate the
comoving correlation lengths at the effective redshifts of our survey
slices we use the photometric redshift distributions presented in
Section~\ref{sec:phot-redsh-pbzk}.

Given a redshift interval $z_1,z_2$ and a redshift distribution
$dN/dz$ we define the effective redshift in the usual way, namely,
$z_{\rm eff}$ is defined as 

\begin{equation}
z_{\rm eff} ={\int_{z_1}^{z_2}z(dN/dz)dz/{\int_{z_1}^{z_2}(dN/dz)dz}}. 
\end{equation}

Using these redshift distributions together with the fitted
correlation amplitudes in presented in
Sections~\ref{sec:galaxies-stars} and ~\ref{sec:galaxy-clustering-at}
we can derive the comoving correlation lengths $r_0$ of each galaxy
population at their effective redshifts using the usual
\cite{1953Apj...117..134,Peebles:1980p5506} inversion. We assume that $r_0$ does not
change over the redshift interval probed. 

It is clear that our use of photometric redshifts introduces an
additional uncertainty in $r_0$. We attempted to estimate this
uncertainty by using the probability distribution
functions associated with each photometric redshift to compute an
ensemble of $r_0$ values, each estimated with a different $n(z)$. The
resulting error in $r_0$ from these many realisations is actually
quite small, $\sim 0.02$ for the $pBzK$ population. Of course,
systematic errors in the photometric redshifts could well be much
higher than this. Figure 9. in \citeauthor{Ilbert:2009p4457} shows the
$1\sigma$ error in the photometric redshifts as a function of
magnitude and redshift. Although all galaxy types are combined here,
we can see that the approximate $1\sigma$ error in the photometric
redshifts between $1<z<2$ is $\sim0.1$.  Our estimate of the
correlation length is primarily sensitive to the median redshift and
the width of the correlation length. An error $\sim0.1$ translates
into an error of $\sim 0.1$ in $r_0$. We conclude that, for our $pBzK$
and $sBzK$ measurements, the dominant source of uncertainty in
our measurements of $r_0$  comes
from our errors on $w$.

We note that previous investigations of the correlation of passive
galaxies always assumed a fixed $\gamma=1.8$; from
Figure~\ref{fig:w_slices} it is clear that our slope is much
steeper. These surveys, however, fitted over a smaller range of
angular scales and therefore could not make an accurate determination
of the slope for the $pBzK$ population. In all cases we fit for both
$\gamma$ and $A_w$.

Our spatial correlation amplitudes for $pBzK$ and $sBzK$ galaxies are
summarised in Table~\ref{tab:fit_r0}. Because of the degeneracy
between $r_0$ and $\gamma$ we also quote clustering measurements as
$r_0^{\gamma/1.8}$. These measurements are plotted in
Figure~\ref{fig:r0g_zed}. At lower redshifts, our field galaxy samples
are in good agreement with measurements for optically selected redder
galaxies from the CFHTLS and VVDS surveys
\citep{2006A&A...452..387M,2008A&A...479..321M}. At higher redshifts,
our clustering measurements for $pBzK$ and $sBzK$ galaxies are in
approximate agreement with the measurements of
\cite{Blanc:2008p3635}. We note that part of the differences with the
measurements of \citeauthor{Blanc:2008p3635} arises from the their
approximation of the redshift distribution of passive $BzK$ galaxies
using simple Gaussian distribution. 

Interestingly, a steep slope $\gamma$ for optically-selected
passive galaxies has already been reported at lower redshift surveys;
for example, \cite{2003MNRAS.344..847M} found that passive galaxies
had a much steeper slope than active galaxies in the 2dF galaxy
redshift survey. 

The highly biased nature of the $pBzK$
  galaxy population indicates that these objects reside in more massive
  dark matter haloes than either the field galaxy population or the
  $sBzK$ population, and} we intend to present a more detailed
discussion of the spatial clustering of each galaxy sample in the
framework of the halo occupation models in a future paper.

\begin{deluxetable*}{ccccccc}
\tablecaption{Spatial Correlation Amplitudes.\label{tab:fit_r0}}
\tablehead{
&\multicolumn{3}{c}{Passive $BzK$ galaxies} & 
\multicolumn{3}{c}{Star-forming $BzK$ galaxies}\\
\colhead{$K_{\rm s}$} &\colhead{$z_{\rm eff}$} & \colhead{$r_0$}&\colhead{$r_0^{\gamma/1.8}$}&
\colhead{$z_{\rm eff}$} & \colhead{$r_0$}&\colhead{$r_0^{\gamma/1.8}$}}
\tablewidth{0pt}
\startdata
22.0 &  1.41 &  $4.55 \pm  0.97$ &  $7.05 \pm  0.51$ & 1.61 & $ 4.69 \pm  0.80$ & $ 4.69 \pm  0.23$\\ 
23.0 &  1.41 &  $3.71 \pm  0.73$ &  $6.18 \pm  0.40$ &  1.71 &  $4.25 \pm  0.43$ &  $4.25 \pm  0.11$\\ 
\enddata
\end{deluxetable*}

\begin{figure}[htb!]
\resizebox{\hsize}{!}{\includegraphics{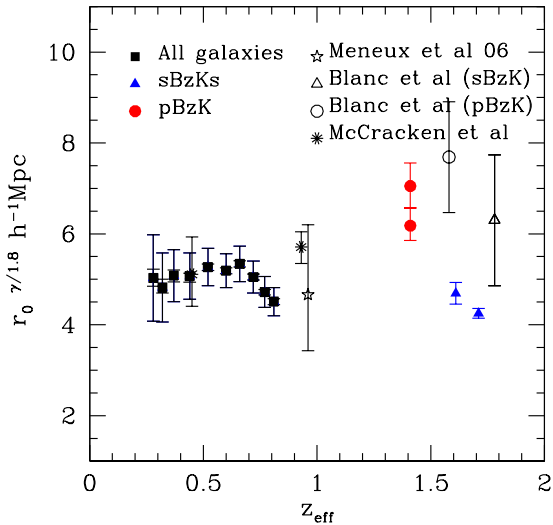}}
\caption{The rescaled comoving correlation length $r_0^{\gamma/1.8}$
as a function of redshift for $K_{\rm s}-$ selected field galaxies (filled
squares), $sBzK$ galaxies (filled triangles) and $pBzK$ galaxies
(filled circles). Also shown are results from lower-redshift optically
selected red galaxies and higher redshift $K_{\rm s}-$ selected samples. }
\label{fig:r0g_zed}
\end{figure}
\nocite{2006A&A...452..387M}

\section{Summary and conclusions}
\label{sec:summary}

We have presented counts, colours and clustering properties for a
large sample of $K-$ selected galaxies in the $2\deg^2$ COSMOS-WIRCam
survey. This represents the largest sample of galaxies to date at this
magnitude limit. By adding deep Subaru $B-$ and $z-$ data we are able
to classify our catalogue into star-forming and quiescent/passive
objects using the selection criterion proposed by
\cite{Daddi:2004p76}. To $K_{\rm s}<23.0$ our catalogues comprises
$143,466$ galaxies, of which $3931$ are classified
as passive galaxies and $25,757$ as star-forming
galaxies. We have also identified a large sample of $13,254$ faint
stars.

Counts of field galaxies and star-forming galaxies change slope at
$K_{\rm s}\sim22$. Our number counts of quiescent galaxies turns over
at $\ks\sim22$, confirming an observation previously made in shallower
surveys \citep{Lane:2007p295}. This effect cannot be explained by
incompleteness in any of our very deep optical bands. Our number
counts of passive, star-forming and field galaxies agree well with
surveys with brighter magnitude limits.

We have compared our counts to objects selected in a semi-analytic
model of galaxy formation. For simple magnitude-limited samples the
\cite{Kitzbichler:2007p3449} model reproduces very well galaxy counts in the
range $16<\ks<20$. However, at fainter magnitudes
\citeauthor{Kitzbichler:2007p3449}'s model predict many more objects than
are observed.

Comparing this model with predictions of passive galaxy counts, we
find that at $20<\kab<20.5$ model counts are below observations by a
factor of 2, whereas at $22.5<\kab<23.0$ model counts are in excess
of observations by around a factor of 1.5. This implies that the
\citeauthor{Kitzbichler:2007p3449} model predicts too many small,
low-mass passively-evolving galaxies and too few large high-mass
passively evolving galaxies at $z\sim1.4$. In these models,
  bulge formation takes place by mergers. At $\ks\sim 22$, passive
  galaxies in the millennium simulation have stellar masses of $\sim
  10^{11}M_\odot$, similar to spectroscopic measurements of passive
  galaxies \citep{Kriek:2006p7151}. This suggests that the difference
  between models and observations is linked to the amount of ``late
  merging'' taking place \citep{DeLucia:2007p7510}. The exact choice of
  the AGN feedback model can also sensitively affect the amount
  star-formation in massive systems
  \citep{DeLucia:2006p9608,Bower:2006p7511}. It is clear that
  observations of the abundance of massive galaxies can now provide
  insight into physical processes occurring in galaxies at intermediate
  redshifts. For the time being it remains a challenge for these models to reproduce both
  these observations at high redshift and lower-redshift reference samples.

  Our results complement determinations of the galaxy stellar
  mass function at intermediate redshifts which show that total mass in
  stars formed in semi-analytic models is too low at $z\sim2$ compared
  to models \citep{Fontana:2006p3228}. We note that
  convolution with standard uncertainties of $\sim0.25$ dex in mass
  function estimates at $z\sim2$ can make a significant difference in the mass
  function, as can be see in Figure~14 of \cite{Wang:2008p9639} who show
  detailed comparisons between semi-analytic models and
  observations. The discrepancy between our observations and models
  cannot be explained in this way.

We have cross-matched our catalogue with precise 30-band photometric
redshifts calculated by \citeauthor{Ilbert:2009p4457} and have used
this to derive the redshift distributions for each galaxy population.
At \kab$\sim 22$ our passive galaxies have a redshift distribution
with $z_{\rm {med}}\sim1.4$, in approximate agreement with similar
spectroscopic surveys comprising smaller numbers of objects. 
Most of our $pBzK$ galaxies have $z_p<2.0$, in contrast with the
redshift distribution for $sBzK$ galaxies and for the general field
galaxy population which extend to much higher redshifts at this
magnitude limit. In the redshift range $1<z<3$ at $\ks\sim22$ the
$pBzK$ population represents around $\sim20\%$ of the total number of
galaxies, in contrast to $\sim80\%$ for $sBzK$-selected
galaxies. DRG-selected galaxies remain an important fraction of the
total galaxy population, reaching around $\sim 50\%$ of the total at
$z\sim2$.  Our work confirms that most galaxies satisfying the
passive-$BzK$ selection criteria lie in a narrower redshift than
either $sBzK$- or DRG-selected objects. Interestingly, a few a passive
$BzK$ galaxies in our survey have $z_p>2.0$, and it is tempting to
associate these objects with higher-redshift evolved galaxies detected in
spectroscopic surveys \citep{Kriek:2008p9702}.
We  have investigated the clustering properties of our catalogues for
which the $2\deg^2$ field of view of the COSMOS survey
provides a unique probe of the distant universe. Our stellar
correlation function is zero at all angular scales to $K_{\rm s}\sim23$
demonstrating the photometric homogeneity and stability of our
catalogues. For a $K_{\rm s}-$ selected samples, the clustering amplitude
declines monotonically toward fainter magnitudes. However, the
slope of the best-fitting angular correlation function becomes
progressively shallower at fainter magnitudes, an effect already seen
in the COSMOS optical catalogues.

At the faintest magnitude slices, the field galaxy population
(all objects with $18.0<\ks<23.0$) is only slightly more
clustered than the underlying dark matter distribution, indicating
that $K_{\rm s}-$ selected samples are excellent tracers of the
underlying mass. On the other hand, star-forming and passive galaxy
candidates are more clustered than the field galaxy population. At
arcminute scales and smaller the passive BzK population is strongly
biased with respect to the dark matter distribution with bias values
of 2.5 and higher, depending on scale. 

Using our photometric redshift distributions, we have derived the
comoving correlation length $r_0$ for each galaxy class. Fitting
simultaneously for slope and amplitude we find a comoving correlation
length $r_0^{\gamma/1.8}$of $\sim7 h^{-1}$~Mpc for the passive $BzK$
population and $\sim 5 h^{-1}$~Mpc for the star-forming $BzK$ galaxies
at $K_{\rm s}<22$ . Our field galaxy clustering amplitudes are in
approximate agreement with optically-selected red galaxies at lower
redshifts. 

High bias values are consistent with a picture in which
$pBzK$ galaxies inhabit relatively massive dark matter haloes on order
of $\sim10^{12}$M$_\odot$, compared to the $sBzK$ and field galaxy population. We will
return to this point in future papers, where will interpret these
measurements in terms of the halo model.

Measuring spectroscopic redshifts for a major fraction of 3000
$pBzK$ galaxies in the COSMOS field one will have to wait for the
advent of large-throughput, wide-field near-infrared ($J$-band)
spectrographs on 8-10m class telescopes, such as FMOS at Subaru
(Kimura et al. 2006). Smaller field, cryogenic, multi-object
spectrographs such as MOIRCS at Subaru \citep{Ichikawa:2006p4968},
EMIR at the GTC telescope \citep{Garzon:2006p4837}, and Lucifer at the
LBT \citep{Mandel:2006p4846} should prove effective in producing high
S/N spectra for a relatively small fraction of the $pBzK$ galaxies in
the COSMOS field.

Since this paper was prepared, additional COSMOS-WIRCam $K_{\rm s}$ data
observations have been taken which will increase the total exposure
time by $\sim30\%$. In addition, new $H$- observations have also
been made. Both these data products will be made publicly available
in around one year from the publication of this article. In the longer
term, the COSMOS field will be observed as part of the UltraVISTA deep
near-infrared survey which will provide extremely deep $JHK$ observations
over the central part of the field.

\section{Acknowledgements}
\label{sec:acknowledgement}

This work is based in part on data products produced at TERAPIX at the
Institut d'Astrophysique de Paris. H.J.~McC acknowledges the use of
TERAPIX computing facilities and the hospitality of the IfA, Honolulu,
where this paper was finished. M. L. Kilbinger is acknowledged for
help with dark matter models in Section 6 and N. V. Asari for the
stacking analysis in Section 4. This research has made use of the
VizieR catalogue access tool provided by the CDS, Strasbourg,
France. This research was supported by ANR grant
``ANR-07-BLAN-0228''. ED and CM also acknowledge support from
``ANR-08-JCJC-0008''.  JPK acknowledges support from the CNRS. We
thank the referee for an extensive commentary on a earlier version of
this paper.

\bibliographystyle{apj}

\end{document}